\documentstyle[preprint,aps,epsf,floats,color]{revtex}

\newcommand{\nn}{\nonumber}

\newcommand{\alpS}{\alpha_{S}}
\newcommand{\alpU}{\alpha_{U}}

\def\bmag#1{{|{\mathbf #1}|}}

\def\Dsl{\hbox{/\kern-.6000em D}} 

\def\vev#1{\left\langle{#1}\right\rangle}
\def\dsl{\,\raise.15ex\hbox{/}\mkern-13.5mu D}
\def\bsigma{\mbox{\boldmath $\sigma$}}
\def\lqcd{\Lambda_{\rm QCD}}
\def\psip#1{\psi_{\mathbf{#1}}}
\def\chip#1{\chi_{\mathbf{#1}}}
\def\bsigma{\mbox{\boldmath $\sigma$}}

\def\ltap{\ \raise.3ex\hbox{$<$\kern-.75em\lower1ex\hbox{$\sim$}}\ }
\def\gtap{\ \raise.3ex\hbox{$>$\kern-.75em\lower1ex\hbox{$\sim$}}\ }
\def\OMIT#1{}

\def\tb#1{\textcolor{blue}{#1}}
\def\tbb#1{\textcolor{black}{#1}}

\def\msb{{\overline{\rm MS}}}
\def\ams#1{\mbox{${\overline\alpha_s^{\,#1}}$}}

\begin{document}
\setlength\baselineskip{20pt}

\preprint{\vbox{\tighten  \hbox{UCSD/PTH 00-25} \hbox{MPI-PhT/2001-04}
\hbox{hep-ph/0102257}
}}

\title{The Running Coulomb Potential and Lamb Shift in QCD}

\author{ Andre~H.~Hoang\thanks{ahoang@mppmu.mpg.de} \\[4pt]} 

\address{\tighten
Max-Planck-Institut f\"ur Physik\\
(Werner-Heisenberg-Institut)\\
F\"ohringer Ring 6, 80805 M\"unchen, Germany}

\author{Aneesh V. Manohar\thanks{amanohar@ucsd.edu}, Iain W.\
Stewart\thanks{iain@schwinger.ucsd.edu} \\[4pt]} 
\address{\tighten Department of Physics, University of California at San
Diego,\\[2pt] 9500 Gilman Drive, La Jolla, CA 92093, USA }

\maketitle

{\tighten
\begin{abstract}

The QCD $\beta$-function and the anomalous dimensions for the Coulomb
potential and the static potential first differ at three loop order. We evaluate
the three loop ultrasoft anomalous dimension for the Coulomb potential and give
the complete three loop running. Using this result, we calculate the leading
logarithmic Lamb shift for a heavy quark-antiquark bound state, which
includes all contributions to the 
binding energies of the form $m\,\alpha_s^4 (\alpha_s\ln \alpha_s)^k$, $k\ge
0$.

\end{abstract}
\pacs{12.39.Hg,11.10.St,12.38.Bx}
}

\section{Introduction}\label{sec:intro}

In this paper, we construct the three-loop anomalous dimension for the Coulomb
potential in non-relativistic QCD (NRQCD)~\cite{nrqcd,pNRQCD}. The formalism we
use was developed in Refs.~\cite{LMR,amis,amis2,amis3} and will be referred to
as vNRQCD, an effective theory for heavy non-relativistic quark-antiquark
pairs. Part of our computation is related to the running of the static
potential~\cite{PSstat}, however effects associated with motion of the quarks do
play an important role. Our final result for the Coulomb potential differs from
the static potential at terms beyond those with a single logarithm (i.e. starting
at four loops). Combining our Coulomb potential running with previous results
for the running of the $1/m$ and $1/m^2$ potentials~\cite{amis,amis3} allows us
to compute the next-to-next-to-leading logarithmic (NNLL) corrections in the
perturbative energy of a heavy $Q\bar Q$ bound state, which includes the sum of
terms $m\, \alpha_s^4 (\alpha_s\ln \alpha_s)^k$, $k\ge 0$, where $m$ is the
heavy quark pole mass. This contribution is the QCD analog of the QED
$\alpha^5\ln\alpha$ Lamb shift computed by Bethe. In QED, the series $m\,
\alpha^4 (\alpha \ln \alpha)^k$ terminates after the $k=1$ term~\cite{amis4}. In
QCD, there is an infinite series due to the running QCD coupling as well as the
presence of non-trivial anomalous dimensions for other QCD operators.  The
results presented here also contribute to the NNLL prediction for the cross
section for $e^+e^-\to t\bar t$ near threshold~\cite{hmst}.  Implications for
$b\bar b$ sum rules will be addressed in a future publication.

The expansion parameter of the effective theory is the quark velocity $v$. A
quark has a momentum of order $mv$ and an energy of order $mv^2$.  We assume
that $m$ is large enough that $mv^2\gg \Lambda_{\rm QCD}$ and a perturbative
description of the bound state as a Coulombic system is valid. For a Coulombic
bound state, $\alpha_s$ is of order $v$ and contributions suppressed by both $v$
and $\alpha_s$ are of the same order.  It is useful to distinguish between
powers of $\alpha_s$ and $v$ when carrying out the matching and when evolving
couplings and operators in the effective theory, and to only take $v\sim
\alpha_s$ for the power counting of bound state matrix elements. In the
effective theory, the quark-antiquark potentials appear as four-quark
operators~\cite{pNRQCD}.  A potential of the form $\alpha_s^r/\bmag{k}^s$ is of
order $\alpha_s^r v^{1-s}$, where ${\bf k}$ is the fermion momentum transfer.
With this power counting the time-ordered product of a $v^a$ and $v^b$ potential
is of order $v^{a+b}$. Up to next-to-next-to-leading order (NNLO) the heavy
quark potential has contributions
\begin{eqnarray} \label{Vpc} 
  V  & \sim & \bigg[\: {\alpha_s \over v} \: \bigg] \ +\ 
  \bigg[\: { \alpha_s^2 \over v} \: \bigg] \ + \ \bigg[\: { \alpha_s^3 \over v} 
  + {\alpha_s^2 v^0 } + {\alpha_s v } \: \bigg] \ +  \ldots \,. 
\end{eqnarray}
The order $\alpha_s/v \sim 1$ term in Eq.~(\ref{Vpc}) is the
Coulomb potential generated at tree-level. The next-to-leading order (NLO) term
is the one-loop correction to the Coulomb potential. The NNLO terms are the
two-loop correction to the Coulomb potential, the one-loop value for the $1/(m
\bmag{k})$ potential, and the tree-level contribution to the order
$\bmag{k}^0/m^2$ potential. Writing the $Q\bar Q$ energy as $E = 2 m + \Delta E$,
the terms in Eq.~(\ref{Vpc}) generate contributions of the following order in the
binding energy:
\begin{eqnarray} \label{Epc} 
  {\Delta E} & \sim & \Big[ m\alpha_s^2 \Big] + \Big[ m\alpha_s^3 \Big]  
  + \Big[ m\alpha_s^4 \Big] + \ldots \,. 
\end{eqnarray}

In Eqs.~(\ref{Vpc}) and (\ref{Epc}) the expansion has been performed at the
scale $\mu=m$ so the coupling constants are $\alpha_s = \alpha_s(m)$. A typical
perturbative expansion contains logarithms of $\mu$ divided by the various
physical scales in the bound state. If the logarithms are large, fixed order
perturbation theory breaks down, and one finds a large residual $\mu$
dependence. One can minimize the logarithms by setting $\mu$ to a value
appropriate to the dynamics of the non-relativistic system.  This is
accomplished by summing large logarithms using the renormalization group, and
using renormalization group improved perturbation theory. For $Q\bar Q$ bound
states, the large logarithms are logarithms of $v\sim \alpha_s$, and can be
summed using the velocity renormalization group (VRG)~\cite{LMR}.  For the
binding energy this gives the expansion
\begin{eqnarray} \label{Ell}
  \Delta E &=& \Delta E^{LL}+ \Delta E^{NLL} + \Delta E^{NNLL} + \ldots \,,\\
 &\sim & \bigg[m \sum_{k=0}^\infty \alpha_s^{k+2}(\ln\alpha_s)^k\bigg] 
 + \bigg[m \sum_{k=0}^\infty \alpha_s^{k+3}(\ln\alpha_s)^k \bigg] 
 + \bigg[m \sum_{k=0}^\infty \alpha_s^{k+4}(\ln\alpha_s)^k \bigg] 
 + \ldots \,,\nn
\end{eqnarray}
where the terms are the leading log (LL), next-to-leading log (NLL), and
next-to-next-to-leading log (NNLL) results respectively.

In the VRG, one uses a subtraction velocity $\nu$ that is evolved from $1$ to
$v$.  This simultaneously lowers the momentum cutoff scale $\mu_S=m\nu$ from $m$
to $mv$ and the energy cutoff scale $\mu_U=m\nu^2$ from $m$ to $mv^2$. The VRG
properly accounts for the coupling between energy and momentum caused by the
equations of motion for the non-relativistic quarks. QED provides a highly
non-trivial check of the VRG method. In Ref.~\cite{amis4} it was used to
correctly reproduce terms in the subleading series of logarithms, including the
$\alpha^3\ln^2\alpha$ corrections to the ortho and para-positronium decay rates,
the $\alpha^7\ln^2\alpha$ hyperfine splittings for Hydrogen and positronium, and
the $\alpha^8 \ln^3 \alpha$ Lamb shift for Hydrogen. The difference between the
VRG, which involves the evolution of the momentum and the energy scale in a
single step, and a conventional two stage renormalization group treatment, $m\to
mv\to mv^2$, was examined in Ref.~\cite{mss1}.

In section~\ref{sec:pots} we review the definition of potentials for non-static
heavy quarks in the effective theory.  In section \ref{sec:static} we compare
these potentials with the Wilson loop definition which is appropriate for static
quarks.  In section~\ref{sec:coulomb} we rederive the leading-logarithmic (LL)
and next-to-leading-logarithmic (NLL) results for the $Q\bar Q$ binding energy
using the effective theory and discuss the two-loop matching for the Coulomb
potential using the results in Refs.~\cite{Peter,Schroeder}.  We also discuss
some subtleties in the correspondence between diagrams in the static theory and
soft diagrams in the effective theory. In section~\ref{sec:threeloop} we compute
the three-loop anomalous dimension for the Wilson coefficient of the Coulomb
potential. Results for the NNLL energy are given in section~\ref{sec:nnll},
followed by conclusions in section~\ref{sec:conclusion}. In Appendix~\ref{UVIR}
we give some technical details on the structure of divergences in the effective
theory, and in Appendix~\ref{Ns} we list some functions that appear in the
energy at NNLL order.

\section{The vNRQCD Potentials}\label{sec:pots}

The effective theory vNRQCD has soft gluons with coupling constant
$\alpS(\nu)$, ultrasoft gluons with coupling constant $\alpU(\nu)$, as well as
quark-antiquark potentials. The potential is the momentum dependent coefficient
of a four-fermion operator:
\begin{eqnarray}\label{Vm}
{\mathcal L} &=&  - \sum_{\mathbf p,p'}  V\left({\bf p},{\bf p^\prime}\right)\: 
  \Big[ {\psip {p^\prime}}^\dagger\: {\psip p}\: 
  {\chip {-p^\prime}}^\dagger\:  {\chip {-p}}{}\Big] \,,
\end{eqnarray}
where spin and color indices are suppressed. The coefficient $V\left({\bf
p},{\bf p^\prime} \right)$ has an expansion in powers of $v$, $ V = V_{(-1)}
+ V_{(0)} + V_{(1)} + \ldots $, where $V_{(-1)}=V_c$ is the Coulomb
potential. For equal mass fermions
\begin{eqnarray} \label{V}
 V_c &=& (T^A \otimes \bar T^A) { \tb{ {\cal V}_c^{(T)}} \over {\mathbf
  k}^2} +  (1 \otimes 1)  { \tb{ {\cal V}_c^{(1)} } \over
  {\mathbf k}^2} ,\nn\\[10pt]
 V_{(0)} &=& (T^A \otimes \bar T^A) { \pi^2 \tb{ {\cal V}_k^{(T)} } \over m\,
 |{\mathbf k}| } + (1 \otimes 1) {\pi^2 \tb{ {\cal V}_k^{(1)} } \over 
 m\,|{\mathbf k}|} \,, \nn \\[10pt]
 V_{(1)} &=& (T^A \otimes \bar T^A) \left[ {\tb{ {\cal
  V}_2^{(T)} } \over m^2 } + { \tb{ {\cal V}_s^{(T)} } {\mathbf
  S}^2 \over m^2}\, +  { \tb{ {\cal V}_r^{(T)} }\: ({\mathbf p^2 +
  p^{\prime 2}}) \over 2\, m^2\, {\mathbf k}^2}  -{i {\tb{ {\cal
  V}_\Lambda^{(T)}} }\, {\mathbf S} \cdot ({\bf p\,^\prime \times p} ) 
  \over m^2 {\mathbf k}^2 } + { \tb{ {\cal V}_t^{(T)} } T({\mathbf k}) \over 
  m^2}\, \right] \nn \\
 && + (1 \otimes 1) \left[{ \tb{ {\cal V}_2^{(1)} } \over m^2}\:+ 
  { \tb{ {\cal V}_s^{(1)} } \over m^2}\: {\mathbf S}^2 \right] , 
\end{eqnarray}
where ${\bf k} = {\bf p'} - {\bf p}$, $\mathbf S = { ({\mathbf \bsigma_1 +
\bsigma_2}) / 2}$, $T({\mathbf k}) = (\delta^{ij} - {3 {\mathbf k}^i{\mathbf
k}^j/{\mathbf k}^2}){ \bsigma_1^i \bsigma_2^j}$.  The Wilson coefficients,
${\cal V}^{(T,1)}$ depend on the subtraction velocity $\nu$. In Eq.~(\ref{V})
the color decomposition $V = (T^A \otimes \bar T^A) { V^{(T)} } + (1 \otimes 1)
{ V^{(1)} } $ has been used and the potential in the color singlet channel is
$V^{(s)}=V^{(1)}-C_F {V}^{(T)}$.  (The Casimirs of the adjoint and fundamental
representations are denoted by $C_A$ and $C_F$, respectively.) At LL order the
running of the coefficients ${\cal V}^{(1,T)}_{2,s,r}$ was computed in
Ref.~\cite{amis} and ${\cal V}^{(1,T)}_{\Lambda,t}$ in Ref.~\cite{amis,chen},
while the NLL order running of ${\cal V}_k^{(1,T)}$ was computed in
Ref.~\cite{amis3}. In this work we compute the running of ${\cal V}_c^{(s)}$ at
NNLL order. This allows the computation of the $Q\bar Q$ energy spectrum at NNLL
order.

In vNRQCD additional potential-like effects are generated by
loops with soft gluons, for which the Feynman rules can be found in
Refs.~\cite{LMR,amis2}. Matrix elements of soft gluon diagrams contribute to
the energy beginning at NLO. In contrast, matrix elements with ultrasoft gluons
start at ${\rm N}^3{\rm LO}$.  The renormalization group improved energies are
obtained by computing the anomalous dimensions for these soft interactions and
the four fermion operators in Eq.~(\ref{V}).

\section{The static potential versus the Coulomb potential}
\label{sec:static}

Parts of our analysis are related to the study of the static limit of QCD which
describes heavy quarks in the $m\to\infty$ limit. We therefore briefly
review the pertinent results which have been derived in this framework.

In position space the static QCD potential is defined as the expectation value
of the Wilson loop operator,
\begin{eqnarray} \label{wilson}
 V_{\rm stat}(r) = \lim_{T\to\infty} {1\over T} \ln \vev{ {\rm Tr}\, P\exp - 
  i g \oint_C A_\mu dx^\mu}, 
\end{eqnarray} 
where $C$ is a rectangle of width $T$ and fixed height $r$.  This potential is
independent of the mass $m$ of the quarks and depends only on $r$.  In QCD
perturbation theory the static potential is known at two-loop
order~\cite{Peter,Schroeder} .  These calculations use static fermion sources
with propagators which are identical to those in Heavy Quark Effective
Theory~\cite{MW}. The exponentiation of the static
potential~\cite{Gatheral,FrenkelTaylor} guarantees that one can avoid dealing
with graphs which have pinch singularities in momentum space. The analysis of
Refs.~\cite{Gatheral,FrenkelTaylor} also gives a prescription for the color
weight factors for different graphs based on the c-web theorem.

\begin{figure}
 \centerline{ \hbox{\epsfxsize=12cm\epsfbox{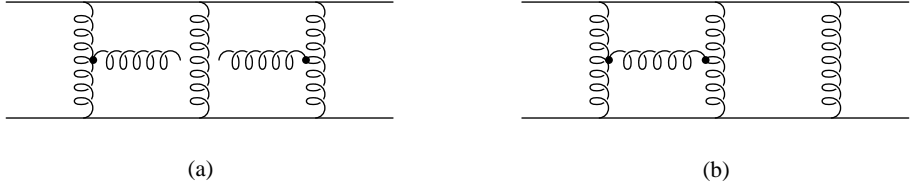} } } \medskip
 {\tighten\caption{Graphs contributing to the three-loop IR divergence of
 the QCD static potential.} \label{fig_ADM} }
\end{figure}

In Ref.~\cite{ADM}, Appelquist, Dine and Muzinich (ADM) pointed out that at
three loops the static potential in Eq.~(\ref{wilson}) has infrared (IR)
divergences from graphs of the form in Fig.~\ref{fig_ADM}a,b.
In the color singlet channel Fig.~\ref{fig_ADM}a has color factor $C_F C_A^2
(C_A-2 C_F)$ while Fig.~\ref{fig_ADM}b is proportional to $C_F^2 C_A^2$.  Taking
into account the exponentiation of $V_{\rm stat}$ using the c-web theorem, the
color singlet contribution to $V_{\rm stat}$ from Fig.~\ref{fig_ADM}(a,b) is
simply Fig.~\ref{fig_ADM}a with the color factor replaced by $C_A^3 C_F$, and is
IR divergent. ADM showed that this IR divergence could be avoided by summing a
class of diagrams---those of Fig.~\ref{fig_ADM} with the addition of an
arbitrary number of gluon rungs.  Summing over these diagrams regulates the IR
divergence by building up Coulombic states for the static quark sources. The
summation gives a factor $\exp\Big([V^{(s)}(r)-V^{(o)}(r)]T\Big)$ for the
propagation of the intermediate color-octet $Q\bar Q$ pair, where $V^{(s)}(r)$
and $V^{(o)}(r)$ are the color-singlet and color-octet potentials. The
exponential factor suppresses long-time propagation of the intermediate
color-octet state, and regulates the IR divergence by introducing an IR cutoff
scale of order $[V^{(s)}(r)-V^{(o)}(r)]\sim \alpha_s/r$.

In Ref.~\cite{static1} the ADM divergence in the static potential was studied by
Brambilla et al.\ using the effective theory pNRQCD~\cite{pNRQCD}. They made the
important observation that along with potential contributions, the definition in
Eq.~(\ref{wilson}) contains contributions from ultrasoft gluons, and the latter
are responsible for the ADM IR divergence. They showed that the ADM IR
divergence in QCD matches with an IR divergence of a pNRQCD graph that describes
the selfenergy of a quark-antiquark system due to an ultrasoft gluon with
momenta $q^\mu \sim \alpha_s/r$.  Therefore, the static potential in pNRQCD can
be defined as a matching coefficient of a four fermion operator, as in
Eq.~(\ref{Vm}), in an infrared safe manner. We will refer to this potential as
the soft-static potential. The ultrasoft pNRQCD graph also has an ultraviolet
divergence.  Brambilla et al.\ computed the coefficient of this divergence and
extracted a new $\ln(\mu)$ contribution to the soft-static potential. In
Ref.~\cite{PSstat} the three-loop anomalous dimension was computed for the
soft-static potential in this framework. In the color singlet channel for scales
$\mu \sim \alpha_s(r)/r$ their solution reads
\begin{eqnarray}\label{psVc}
 V_{\rm stat}(\mu,r) = V_{\rm stat}^{\rm (2 loops)}(r)- 
 \frac{1}{4\pi r} \Bigg[\, \frac{2\pi C_F C_A^3}
 {3\beta_0}\alpha_s^3(r) \ln\bigg(\frac{\alpha_s(r)}{\alpha_s(\mu)}\bigg) 
 \,\Bigg]  \,,
\end{eqnarray}
where the first term is the two loop static potential derived in
Refs.~\cite{Peter,Schroeder}.

\begin{figure}
 \centerline{\epsfxsize=4cm \epsfbox{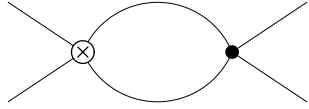}} \medskip
 {\tighten\caption{Loop with an insertion of the $1/(m|{\bf k})$ and 
 $1/{\bf k^2}$ potentials.}
 \label{fig:tproduct}}
\end{figure}

For large but finite $m$ the effective theory for $Q\bar Q$ bound states has an
expansion in $v$.  The quark potential in this case differs from that in the
static case, and in general one cannot obtain the static theory by taking the $m
\to \infty$ limit. To illustrate this consider as an example the loop graph
involving the iteration of a $1/(m\bmag{k})$ and a Coulomb potential as shown in
Fig.~\ref{fig:tproduct}.
In the effective theory for non-static fermions, the fermion propagators give a
factor proportional to $m$.  The final result for the graph is independent of
$m$ and is the same order in $v$ as the Coulomb potential. On the other hand, if
we first take the static limit $m \to \infty$, then $1/(m\bmag{k})\to 0$, and
there is no such graph.  This example illustrates the general result that the
static theory is not obtained as the $m\to\infty$ (or $v\to0$) limit of the
non-static effective theory. Loop integrals can produce factors of $m$ or $1/v$,
and thereby cause mixing between operators which are of different
orders\footnote{\tighten However, we stress that if powers of $\alpha_s$ are
also counted as powers of $v$, then operators which are higher order in $v$
never mix into lower order operators.}  in $v$. Furthermore, the energy and the
momentum transfer in Coulombic states are coupled by the quark equations of
motion. We will see that for this reason the anomalous dimension of the static
and Coulomb potentials differ at three loops. The three-loop matching for the
static and Coulomb potentials can also differ.

\section{The Coulomb potential at one and two loops}\label{sec:coulomb}

In this work dimensional regularization and the $\msb$ scheme will be used.  The
$\msb$ QCD running coupling constant will be denoted by $\ams{}(\mu)$, and is
determined by the solution of the renormalization group equation
\begin{eqnarray}\label{msbar}
 \mu {{\rm d \ams{}(\mu) \over {\rm d} \mu}} &=& 
  \overline \beta \left( \ams{}(\mu) \right)\nn\\
 &=& -2\beta_0 \frac{\ams{2}(\mu)}{4\pi} 
  -2 \beta_1 \frac{\ams{3}(\mu)}{(4\pi)^2}
  -2 \beta_2 \frac{\ams{4}(\mu)}{(4\pi)^3} + \ldots \,.
\end{eqnarray}
In a mass-independent subtraction scheme $\beta_0$ and $\beta_1$ are
scheme-independent. The notation $\ams{[n]} \left(\mu \right)$
will be used to indicate the solution of Eq.~(\ref{msbar}) with coefficients up
to $\beta_{n-1}$ (i.e.\ $n$ loop order) kept in the $\beta$-function.


In the VRG, the soft and ultrasoft subtraction scales $\mu_S$ and $\mu_U$ are
given by $\mu_S=m \nu$ and $\mu_U=m \nu^2$. We define the soft and ultrasoft
anomalous dimensions $\gamma_S$ and $\gamma_U$ as the derivatives with respect
to $\ln\mu_S$ and $\ln \mu_U$, respectively. The derivative with respect to $\ln
\nu$ gives the total anomalous dimension, $\gamma=\gamma_S+2\gamma_U$.  vNRQCD
has three independent but related coupling constants that are relevant for our
calculation: the soft gluon coupling $\alpS(\nu)$, the ultrasoft gluon
coupling $\alpU(\nu)$, and the coefficient of the Coulomb potential ${\cal
V}_c(\nu)$.  The tree-level matching conditions at ($\mu=m \Leftrightarrow
\nu=1$) are
\begin{eqnarray}
 \alpS(1) &=& \alpU (1) = \ams{} (m) \,, \qquad
 {\cal V}_c^{(T)}(1) = 4 \pi \ams{} (m)\,, \qquad 
 {\cal V}_c^{(1)}(1)=0 \, ,
\end{eqnarray} 
and the solutions of the one-loop renormalization group equations for the
coupling constants in the effective theory are~\cite{Gries,LMR}
\begin{eqnarray}\label{loop1}
 \alpS\left( \nu \right) &=& \ams{[1]} (m \nu) \,,\qquad\quad\ \
 \alpU \left( \nu \right) = \ams{[1]} (m \nu^2) \,, \nn\\
 {\cal V}_c^{(T)}(\nu) &=& 4 \pi \ams{[1]}(m \nu) \,,\qquad
 {\cal V}_c^{(1)}(\nu) = 0.
\end{eqnarray}
In deriving the above equations, it has been assumed that any light fermions
have masses much smaller than $mv^2$, so that there are no mass thresholds in
the renormalization group evolution. If there is a mass threshold larger than
$mv^2$ and widely separated from $mv$ and $m$, then it is possible to also
include such effects in the effective theory, see Ref.~\cite{LMR}.

At leading order the Hamiltonian for the color singlet $Q\bar Q$ system is
\begin{eqnarray}
  H_0 = \frac{\bf p^2}{m} + \frac{ \tbb{ {\cal V}_c^{(s)}(\nu)} }{\bf k^2} \,. 
\end{eqnarray}
To minimize large logarithms in higher order matrix elements we run $\nu$ to
the bound state velocity $v_b$, which we define as the solution of the equation
\begin{eqnarray}
 v_b =  \frac{ \tbb{ a_c(\nu=v_b) }}{ n} = 
 \frac{ C_F\: \ams{[1]}(m v_b)}{n} \,,
\end{eqnarray}
where for convenience we have defined 
\begin{eqnarray}
   a_c(\nu)=-\: \frac{ {\cal V}_c^{(s)}(\nu)} {4\pi } \,,
\end{eqnarray}
and $n$ is the principal quantum number.  The LL binding energy is then simply
the eigenvalue of the Schr\"odinger equation, $H_0 |\psi_{n,l}\rangle = \Delta E
|\psi_{n,l}\rangle$ with the LL solution for the Coulomb potential, ${\cal
V}_c^{(s)}(\nu)=-C_F {\cal V}_c^{(T)}(\nu)$ from Eq.~(\ref{loop1}).
Thus,
\begin{mathletters}
\begin{eqnarray} \label{eLL}
  \Delta E^{LL} &=& 
   -\frac{m }{4n^2}\: \big[ \tb{ a_c(\nu) } \big]^2 \\
   &=& - \frac{m}{4\,n^2}\: C_F^2 \,\Big[ \ams{[1]}(m v_b)\Big]^2 
     = -\frac{m v_b^2}{4} \,,
\end{eqnarray}
\end{mathletters}
where in the second line we have evaluated the energy at the low scale $\nu=v_b$.
Higher order corrections to the energy are all evaluated as perturbative matrix
elements with the leading order wavefunctions, $|\psi_{n,l}\rangle$.


Consider how the results in Eq.~(\ref{loop1}) are extended to higher orders.
The graphs for the renormalization of the ultrasoft gluon self coupling have
the same rules as for QCD, and those for the renormalization of the lowest
order soft gluon vertex have the same rules as for HQET. Since the
momenta of soft and ultrasoft gluons are cleanly separated there is no mixing
of scales, so the anomalous dimension for $\alpS$ is independent of $\alpU$ and
vica-versa.  Thus, one expects that in the $\msb$ scheme $\alpS\left( \nu
\right) = \ams{}(m \nu)$ and $\alpU\left( \nu \right) = \ams{}(m \nu^2)$ to all
orders.

However, the coefficient of the Coulomb potential can differ from $4 \pi
\ams{}(m\nu)$ at higher orders. At one-loop, the only order $\alpha_s^2/v$ graph
in the effective theory is the soft diagram~\cite{amis2}\footnote{\tighten Note
that the soft loop includes soft gluons, soft light quarks, as well as soft
ghosts.}
\begin{eqnarray}  \label{eftvc1}
\begin{picture}(60,30)(1,1)
 \epsfxsize=2.0cm \lower16pt \hbox{\epsfbox{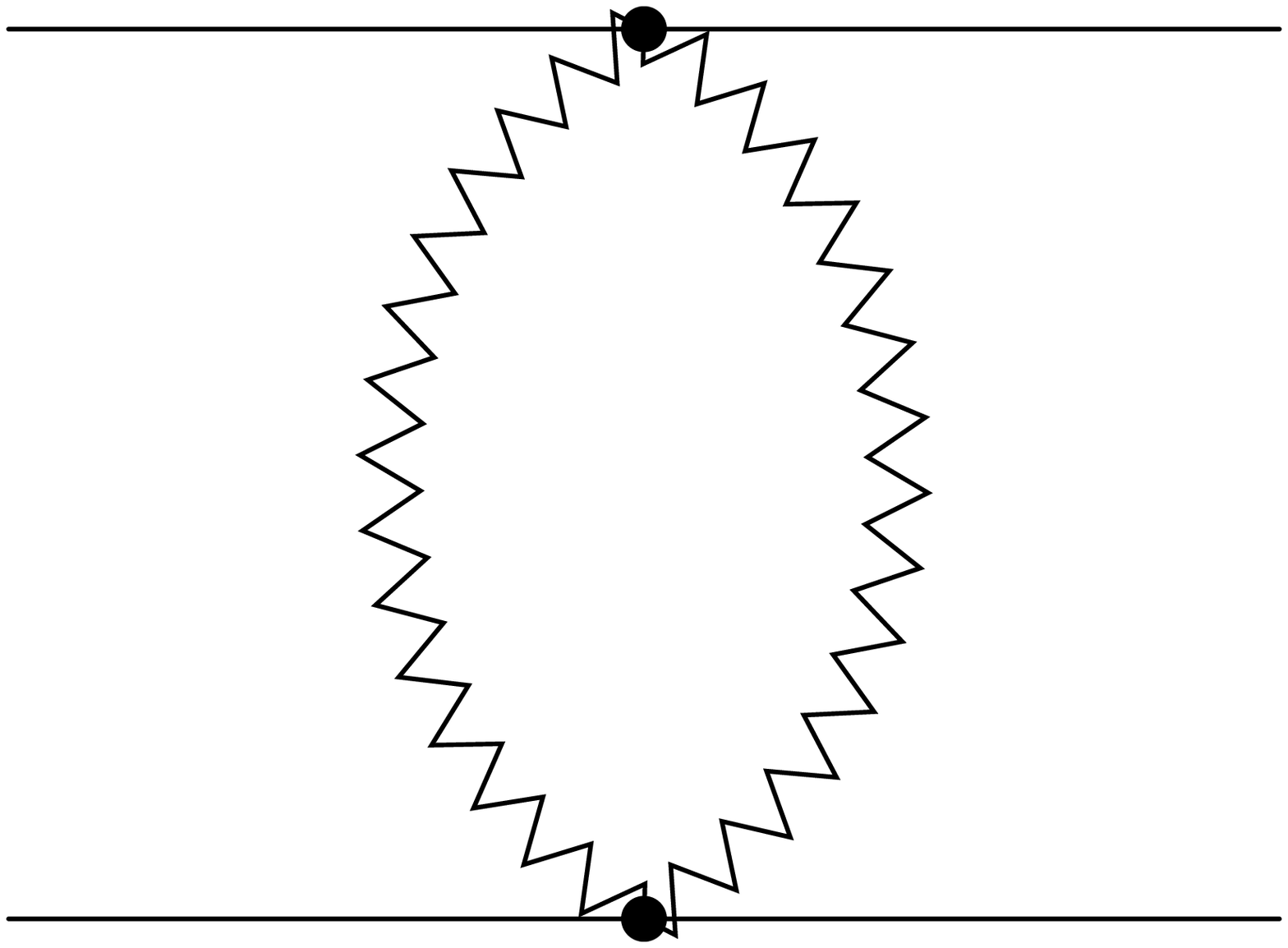}}
\end{picture}
  &=& \frac{-i \mu_S^{2\epsilon}\alpS^2(\nu)}{\bf k^2} (T^A  \otimes \bar T^A) 
  \bigg[ \frac{\beta_0}{\epsilon} + \beta_0
   \ln\Big( \frac{\mu_S^2}{\bf k^2}\Big)  + a_1 \bigg] \,, \\[-15pt]\nn
\end{eqnarray}
where $\beta_0=11/3 C_A-4 T_F n_\ell/3$ and $a_1=31 C_A/9-20 T_F n_\ell/9$ in
the $\msb$ scheme, and $n_\ell$ is the number of light soft quarks. The
divergence in Eq.~(\ref{eftvc1}) is canceled by a counterterm for ${\cal
V}_c^{(T)}$, causing it to run with anomalous dimension
$-2\beta_0\alpS^2(\nu)$. The remaining terms in the soft graph are identical to
the one-loop soft-static potential calculation and also reproduce the set of
$\alpha_s^2/{\bf k^2}$ terms in full QCD with dimensional regularization
parameter $\mu$. The one-loop matching correction to ${\cal V}_c^{(T,1)}(1)$ is
the difference between the full and effective theory diagrams and therefore
vanishes at the matching scale $\mu=\mu_S=m$.

\begin{figure}
 \centerline{ \hbox{\epsfxsize=10cm\epsfbox{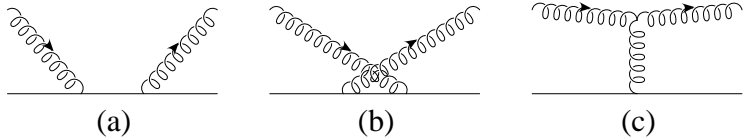} } } 
 {\tighten\caption{Compton scattering graphs that contribute to the soft
 vertex.} 
 \label{fig_compton} }
\end{figure}

\begin{figure}
 \centerline{ \hbox{\epsfxsize=8cm\epsfbox{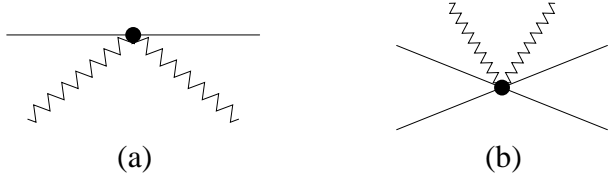} } } 
 {\tighten\caption{Examples of vertices involving soft gluons.} 
 \label{fig_softfr} }
\end{figure}

A correspondence between the soft-static potential calculation and soft order
$1/v$ diagrams is expected to persist at higher orders in $\alpha_s$ as
well. The Feynman rules for the soft vertices are almost identical to the HQET
rules used for soft-static potential calculations. There are a few notable
differences.  In the effective theory it is not necessary to use the
exponentiation theorem~\cite{Gatheral,FrenkelTaylor} to eliminate diagrams with
pinch singularities of the form
\begin{eqnarray}
 \int {\rm d}q^0\ \frac{1}{(q0+i\epsilon)(-q0+i\epsilon)} \,.
\end{eqnarray}
These are automatically removed in the construction of the tree level soft
vertices because the $1/q_0$ factors in the soft Feynman rules do not contain
$i\epsilon$'s, and in evaluating diagrams these poles are ignored.  For
example~\cite{LMR}, from matching the full theory Compton scattering graphs in
Fig.~\ref{fig_compton}
one obtains the soft vertex in Fig.~\ref{fig_softfr}a which is proportional to
$[T^A, T^B]/q_0$, where terms proportional to $\{T^A, T^B\}$ have canceled.  In
the soft-static calculations this cancellation instead takes place at the level
of the box and crossed box graphs, and is guaranteed by exponentiation.  For the
soft-static potential it is known that at higher orders there are contributions
from the $i\pi \delta(q_0)$ terms that originate from the $i\epsilon$'s. It was
exactly this type of contribution that was missed in the two-loop calculation by
Peter~\cite{Peter}, and was correctly identified by
Schr\"oder~\cite{Schroeder}. In the effective theory these delta function
contributions belong to the potential regime~\cite{amis3}, and soft-static graphs
with this type of contribution are reproduced by operators such as the one shown
in Fig.~\ref{fig_softfr}b, where a soft gluon scatters from a
potential. Matching induces these operators to account for the difference
between the full and effective theory graphs for Compton scattering off two
quarks.  Thus, the treatment of $i\epsilon$'s does affect the correspondence
between soft-static and soft graphs. The total contribution of the graphs in the
static theory with $k^\mu\sim mv$ gluons is reproduced in the effective theory
by graphs with soft gluons.

%
%

A real difference between the soft-static and effective theory calculations is
the way in which counterterms are implemented. The soft-static potential is
defined by local HQET-like Feynman rules and all UV divergent contributions from
soft gluons are absorbed into vertex, field, and coupling renormalization.  The
renormalization of the four point function is taken care of by the
renormalization of the two and three point functions.  Renormalization of the
vNRQCD diagrams is quite different because potential gluons are not treated as
degrees of freedom.  The effective theory has graphs with soft gluons and in
addition the four-quark Coulomb potential operator.  The overall divergence in
soft graphs, such as the one in Eq.~(\ref{eftvc1}), are absorbed by ${\cal
V}_c$, while subdivergences are taken care of by counterterms for the soft
vertices and lower order ${\cal V}_c$ counterterms. The sum of unrenormalized
soft-static and purely soft diagrams in the effective theory agree. So if these
were the only considerations, then the description of these effects would be
basically a matter of convenience.  However, divergences associated with
ultrasoft gluons can only be absorbed into ${\cal V}_c$, so at the level that
these gluons contribute in the effective theory it is necessary to adopt the
four quark operator description from the start (i.e. just below the scale $m$).

\begin{figure}
 \centerline{\hbox{\epsfxsize=3cm \epsfbox{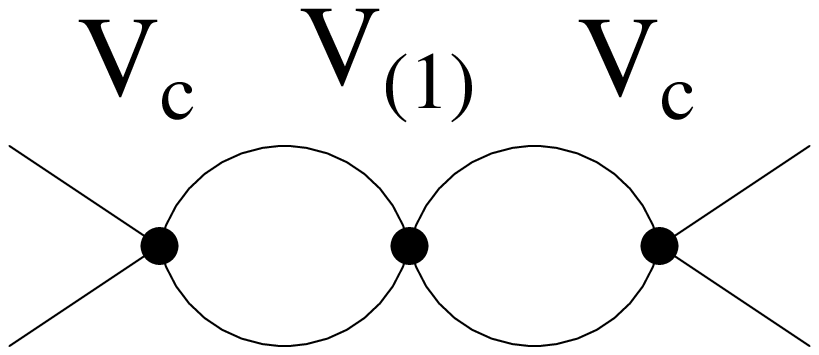}} \qquad 
             \hbox{\epsfxsize=3cm \epsfbox{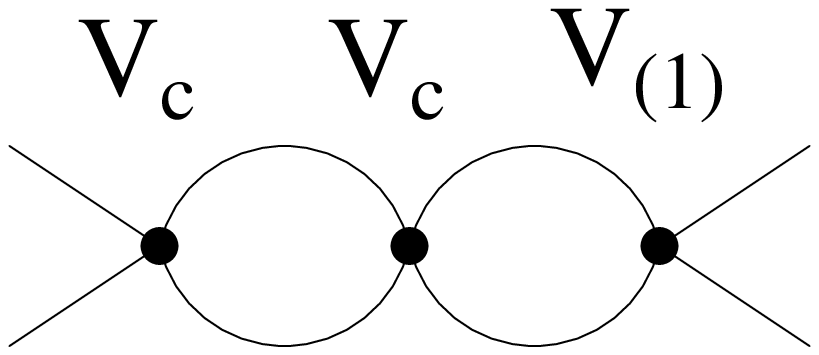}} \qquad
             \hbox{\epsfxsize=3cm \epsfbox{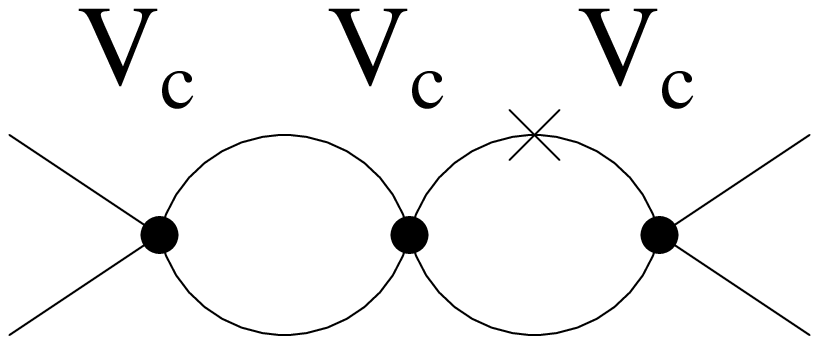}} \qquad
	     \hbox{\epsfxsize=2.3cm \epsfbox{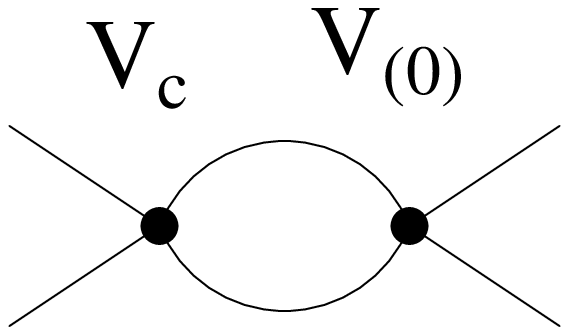}} } \medskip\medskip
{\tighten\caption{Order $\alpha_s^3/v$ diagrams with potential 
 iterations. The $\times$ denotes an insertion of the ${\bf p^4}/8m^3$ 
 relativistic correction to the kinetic term.} 
\label{eft_p2}}
\end{figure}

\begin{figure}
 \centerline{\hbox{\epsfxsize=2.cm \epsfbox{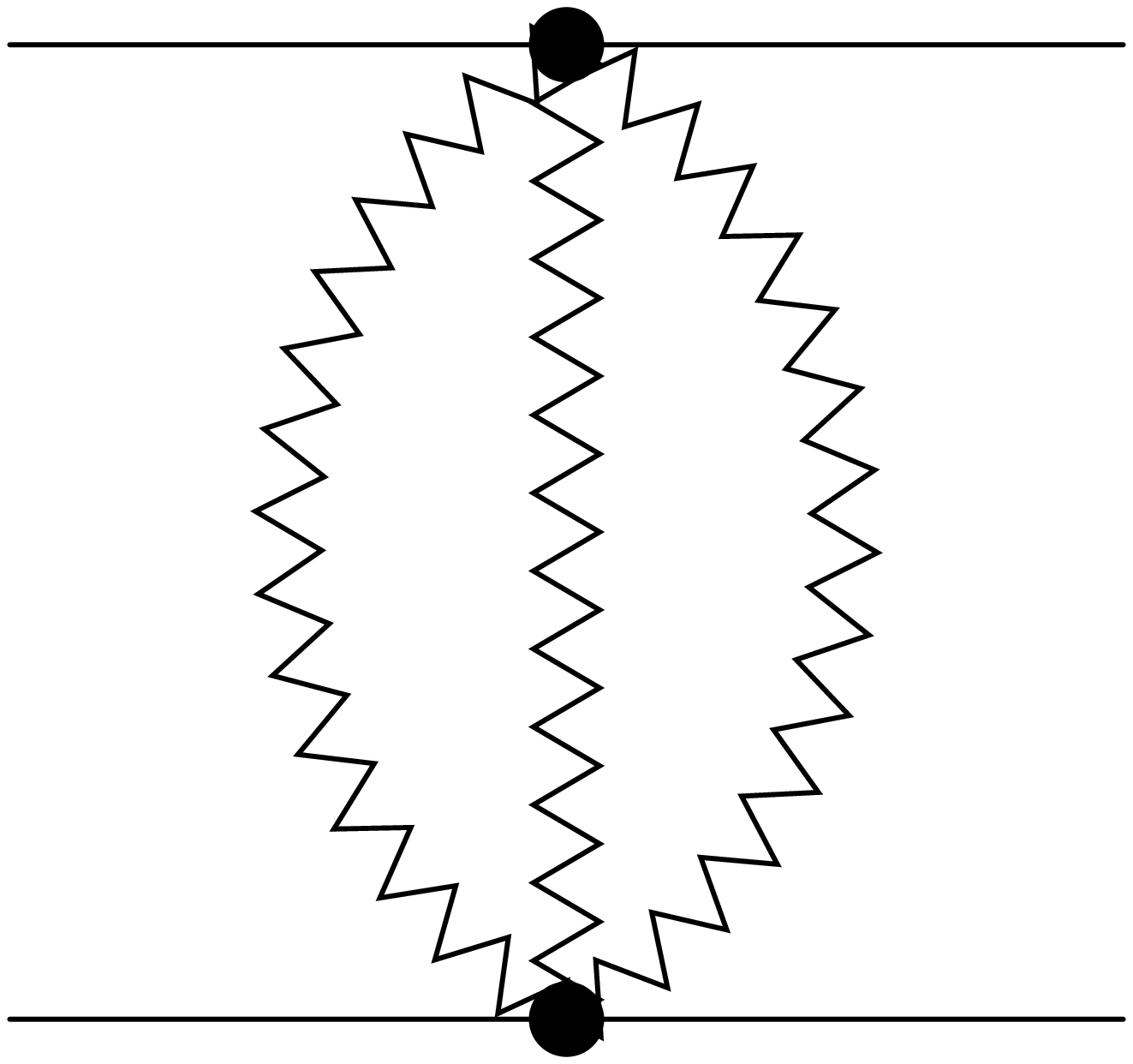}} \quad
             \hbox{\epsfxsize=2.cm \epsfbox{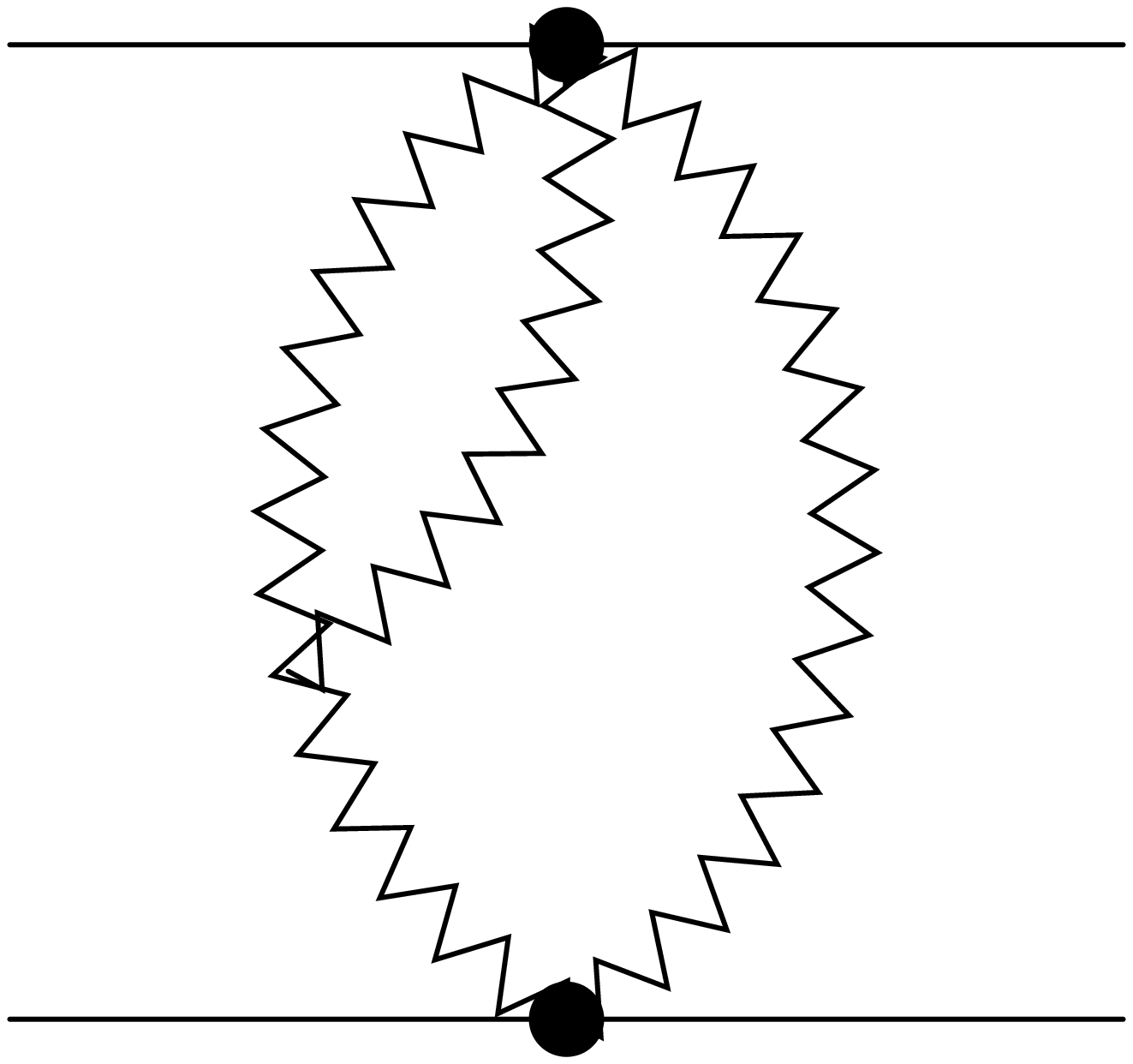}} \quad
             \hbox{\epsfxsize=2.cm \epsfbox{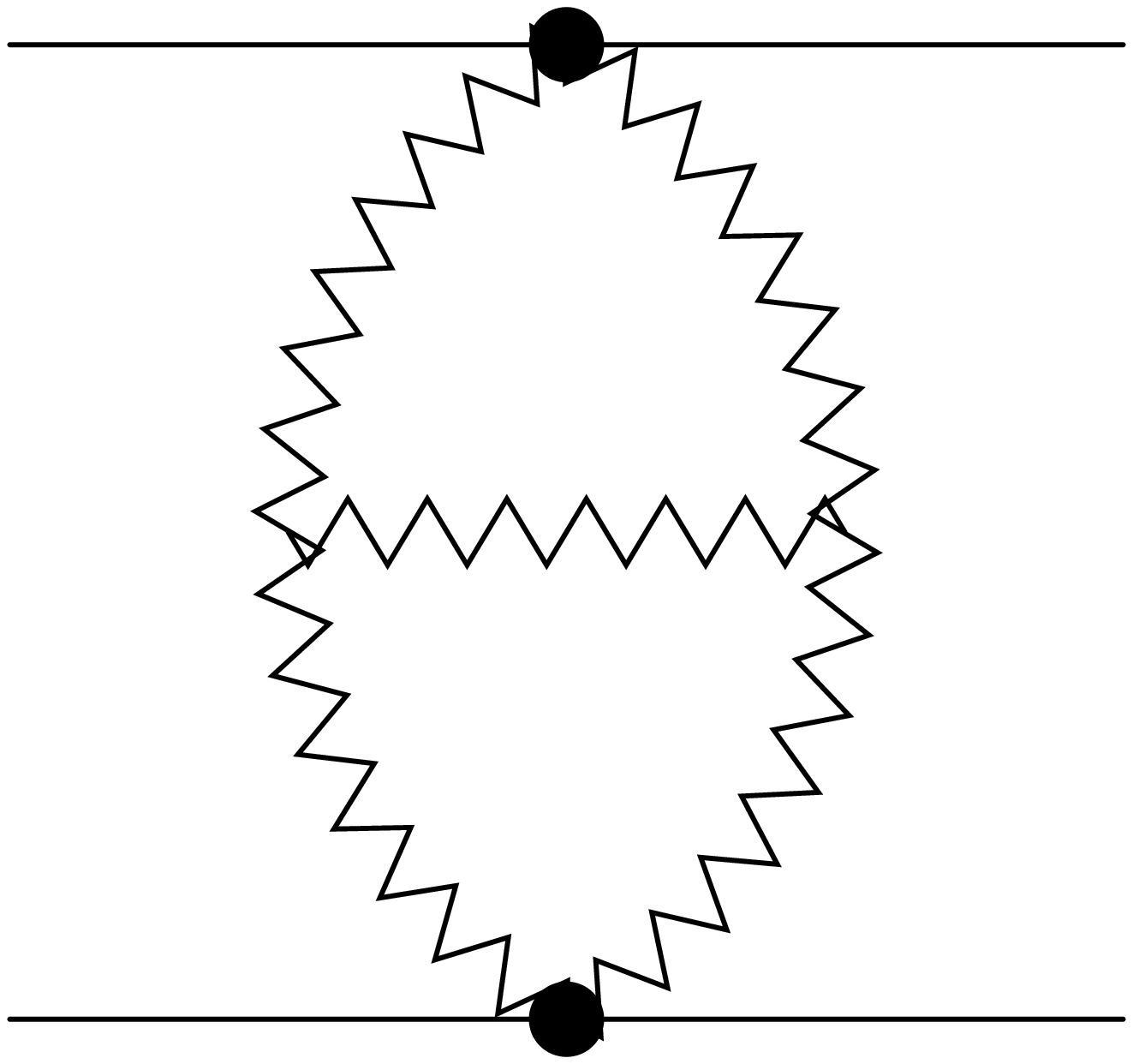}} \quad
           \lower2pt  \hbox{\epsfxsize=2.3cm \epsfbox{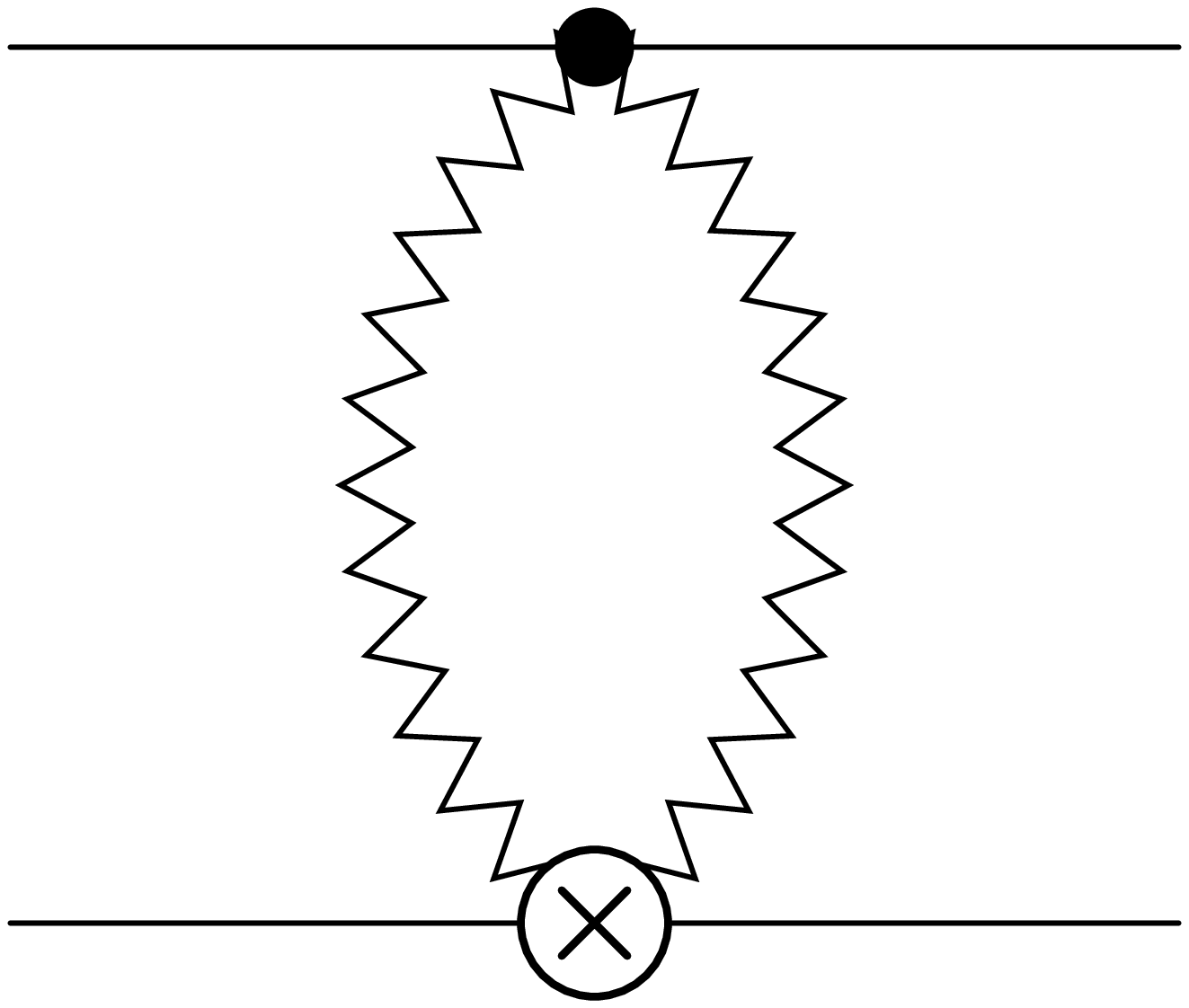}} 
 \raise20pt \hbox{\Large $\ \ \ldots$} } \medskip
 {\tighten\caption{Examples of order $\alpha_s^3/v$ diagrams with soft vertices.
 The vertex with a cross denotes an insertion of a one-loop counterterm. }
 \label{eft_s2}}
\end{figure}

To calculate the NLL energy we also need the two-loop anomalous dimension for
${\cal V}_c$, which is obtained from the renormalization of order $\alpha_s^3/v$
diagrams.  At this order there are effective theory graphs with iterations of
potentials, shown in Fig.~\ref{eft_p2}, and soft effective theory diagrams, as
in Fig.~\ref{eft_s2}.  Graphs with ultrasoft gluons do not contribute at this
order.
The potential diagrams are finite in the ultraviolet and reproduce the Coulombic
singularities in perturbative QCD. The contributions from the soft diagrams can
be determined from the soft-static potential calculations. For the color singlet
channel, the UV divergences in the soft-static two-loop diagrams were calculated
in Ref.~\cite{Fischler} and the constant terms in
Refs.~\cite{Peter,Schroeder}. The sum of unrenormalized soft diagrams has the
form
\begin{eqnarray}  \label{stat1}
  \frac{i \alpS^3(\nu)}{\bf k^2} \frac{C_F}{4\pi}
   \bigg[\frac{\beta_0^2}{\epsilon^2} + \frac{\beta_1\!+\!2\beta_0 a_1}
   {\epsilon} 
   + \frac{2\beta_0^2}{\epsilon}\ln\Big( \frac{\mu_S^2}{\bf k^2}\Big) 
   \!+\! \beta_0^2 \ln^2\Big( \frac{\mu_S^2}{\bf k^2}\Big) \!+\!
   (\beta_1\!+\!2\beta_0 a_1)\ln\Big( \frac{\mu_S^2}{\bf k^2}\Big) 
   + a_2^{(a)} \bigg] \, .
\end{eqnarray}
The effective theory counterterm graphs give
\begin{eqnarray}  \label{eft_ct}
  -\frac{i \alpS^3(\nu)}{\bf k^2} \frac{C_F}{4\pi}
   \bigg[\frac{2\beta_0^2}{\epsilon^2} 
   + \frac{2\beta_0^2}{\epsilon} \ln\Big( \frac{\mu_S^2}{\bf k^2}\Big)
   + \frac{2 a_1 \beta_0}{\epsilon} + a_2^{(b)} \bigg] \,.
\end{eqnarray}
Taking the sum of Eqs.~(\ref{stat1}) and (\ref{eft_ct}) we find that up to
two-loops the counterterm for the color singlet Coulomb potential has the form
\begin{eqnarray} 
 Z_c = 1- \frac{\alpS(\nu)\beta_0}{4\pi\epsilon}  + 
 \frac{\alpS^2(\nu)}{(4\pi)^2} \bigg[ 
 \frac{\beta_0^2}{\epsilon^2} - \frac{\beta_1}{\epsilon} \bigg]
  \,,
\end{eqnarray}
where $\beta_1 = 34 C_A^2/3-4 C_F T_F n_\ell -20 C_A T_F n_\ell/3$.  The
$\alpha_s^2/\epsilon$ divergence is proportional to $\beta_1$, so the two-loop
anomalous dimension for ${\cal V}_c^{(s)}$ is determined by the two-loop
$\overline{\rm MS}$ $\beta$-function, and the NLL coefficient of the singlet
Coulomb potential is ${\cal V}_c^{(s)}(\nu) = 4\pi\ams{[2]}(m \nu)$.

The energy at NLL order involves including the NLL coefficient for the Coulomb
potential, ${\cal V}_c^{(s)}(\nu)$ in Eq.~(\ref{eLL}), and calculating the
matrix element of the one-loop order $1/v$ soft diagram between Coulombic states
[$a_c(\nu)\equiv -{\cal V}_c^{(s)}(\nu)/4\pi$]
\begin{eqnarray} \label{s1}
 i \: \Bigg\langle \begin{picture}(55,20)(-10,1)
  \epsfxsize=1.2cm \lower9pt \hbox{\epsfbox{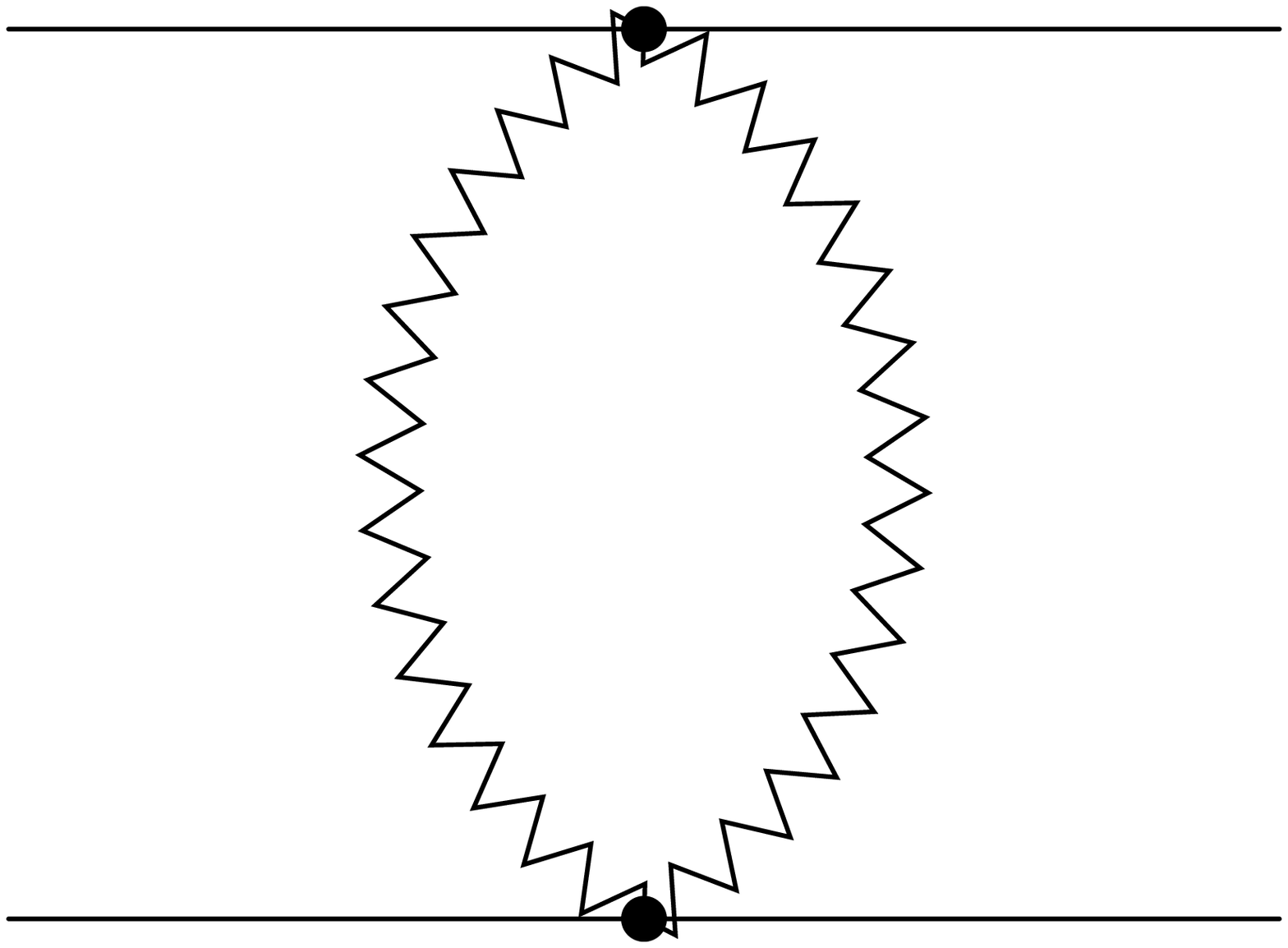}}
 \end{picture}\Bigg\rangle
&=& -C_F\: {\alpS^2(\nu)} \left\langle \frac{1}{\bf k^2}
 \bigg[ a_1 + \beta_0 \ln\Big( \frac{\mu_S^2}{\bf k^2} \Big) \bigg]
\right\rangle 
  \\[5pt]
&=& - \frac{m\,C_F\,\alpS^2(\nu) \tb{ a_c(\nu)} }{8\pi\,n^2}\: 
  \Bigg\{ {a_1} + 2 \beta_0\,\bigg[ \ln\Big(\frac{n\: \nu}{\tb{ a_c(\nu) }}\Big) 
  +\psi(n+l+1)+\gamma_E\bigg] \Bigg\} \,. \nn
\end{eqnarray}
As expected, at the low scale $\nu\simeq v_b$, there are no large logarithms in
the matrix element. Combining Eq.~(\ref{s1}) with Eq.~(\ref{eLL}) gives the
energy valid at NLL order,
\begin{eqnarray} \label{eNLL}
  \Delta E^{LL}+\Delta E^{NLL} &=&  -\frac{m}{4n^2}\: \big[\tb{ a_c(v_b) }\big]^2
 -\frac{m\,C_F\,\alpS^2(v_b) \tb{ a_c(v_b) }}{8\pi\,n^2}\: 
 \bigg\{  2\beta_0 \Big[ \psi(n+l+1)+\gamma_E \Big] + {a_1} \bigg\} \,.\nn\\
\end{eqnarray}

In the next section, the three-loop running of the Coulomb potential will be
derived. We therefore need the two loop matching condition, and so consider the
finite parts for the two loop graphs. The sum of renormalized soft diagrams in
Fig.~\ref{eft_s2} is
\begin{eqnarray}  \label{eftvc2}
  \frac{i \alpS^3(\nu)}{\bf k^2} \frac{C_F}{4\pi} \bigg[\beta_0^2
   \ln^2\Big( \frac{\mu_S^2}{\bf k^2}\Big) +
   (\beta_1+2\beta_0 a_1)\ln\Big( \frac{\mu_S^2}{\bf k^2}\Big) 
   + a_2 \bigg] \,,
\end{eqnarray}
where from Ref.~\cite{Schroeder} the sum of constants in Eqs.~(\ref{stat1}) and
(\ref{eft_ct}) is $a_2=a_2^{(a)}+a_2^{(b)} = 456.75-66.354 n_\ell+1.235
n_\ell^2$ for $n_\ell$ light flavors.  The matching coefficient for ${\cal V}_c$
at the scale $m$ is given by the difference between the $1/{\bf k}^2$ terms in
the $Q\bar Q$ scattering amplitude in the full and effective theories.  It is
convenient to analyze the two loop result in the full theory by using regions in
the threshold expansion~\cite{Beneke}. The soft region exactly reproduces the
result from the soft graphs. Furthermore, the potential region exactly
reproduces the results for the potential graphs in Fig.~\ref{eft_p2}. Thus, the
matching correction for ${\cal V}_c(1)$ is also zero at two-loops. In general, a
non-zero matching correction appears when there is a full theory contribution
from an off-shell region such as the hard regime or when UV divergences appear
in the effective theory graphs.\footnote{\tighten An example where UV
divergences in the effective theory affect the matching is the two-loop
coefficient for the production current.}  In the full theory at two loops there
are no contributions proportional to $1/{\bf k}^2$ from off-shell regions. The
soft effective theory graphs are UV divergent, however these divergences are in
one-to-one correspondence with UV divergences in the full or static theory.
Finally, the graphs with iterations of potentials are UV finite.


\section{Three-loop running of ${\cal V}_{\lowercase{c}}$}\label{sec:threeloop}

\begin{figure}
 \centerline{
 \hbox{\epsfxsize=3.5cm\epsfbox{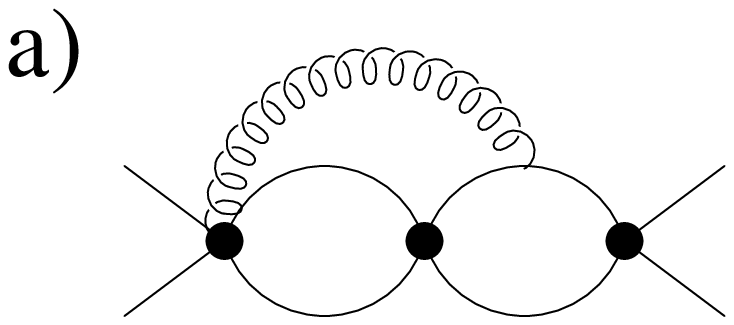}} \quad
 \hbox{\epsfxsize=3.5cm\epsfbox{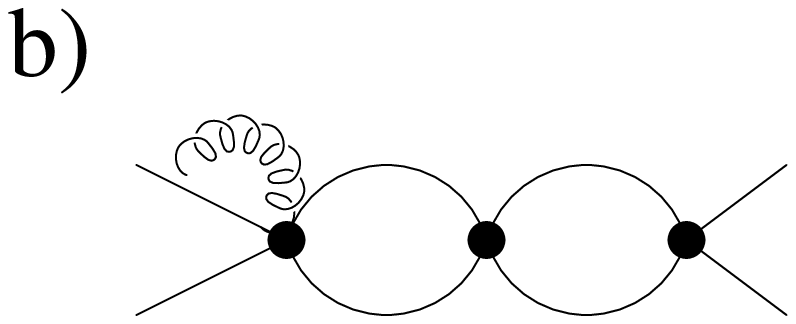}} \quad 
 \hbox{\epsfxsize=3.5cm\epsfbox{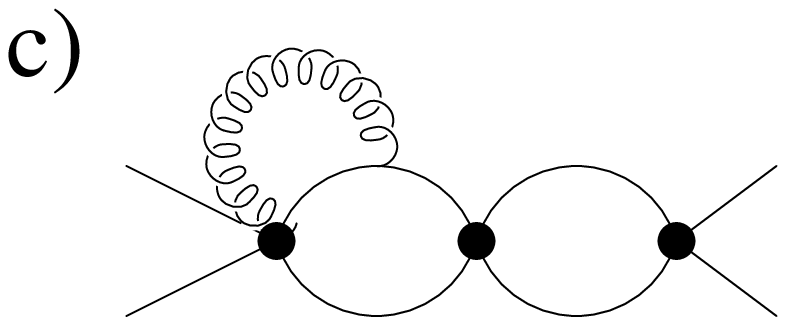}} } \medskip\medskip
 \centerline{
 \hbox{\epsfxsize=3.5cm\epsfbox{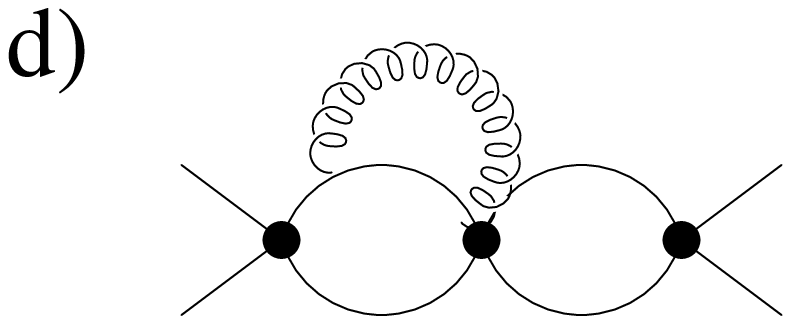}} \quad
 \hbox{\epsfxsize=3.5cm\epsfbox{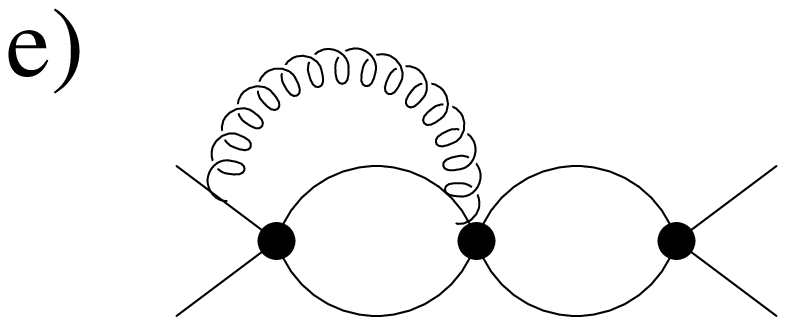}} \quad
 \hbox{\epsfxsize=3.5cm\epsfbox{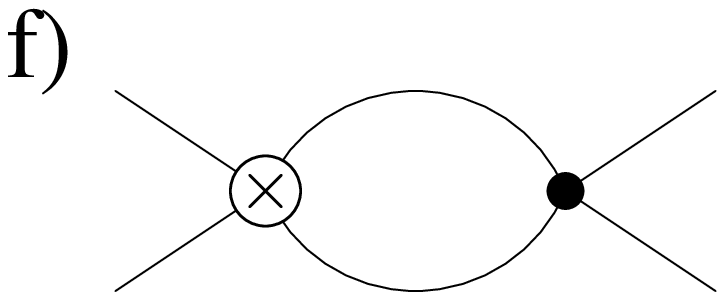}}\quad
 \hbox{\epsfxsize=3.5cm\epsfbox{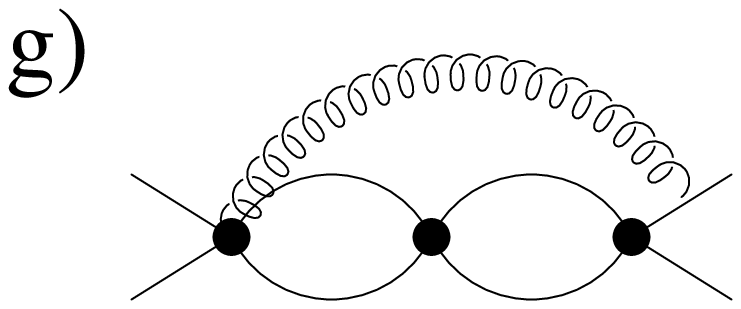}}  
 } \medskip\medskip
 {\tighten \caption{Graphs with ultrasoft gluons which do not contribute to the
 running of the Coulomb potential. The divergences in a)-e) are 
 canceled by graph f) which has an insertion of the corresponding ${\cal V}_k$ 
 counterterm(s) denoted by $\otimes$. Graph g) is UV finite.}
 \label{fig_us1} }
\end{figure}

\begin{figure}
 \centerline{
 \hbox{ \epsfxsize=4.2cm\epsfbox{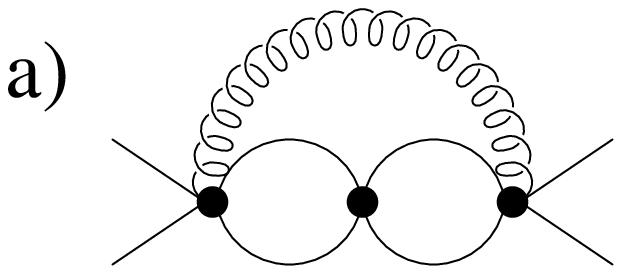}} \qquad 
 \hbox{\epsfxsize=4.2cm\epsfbox{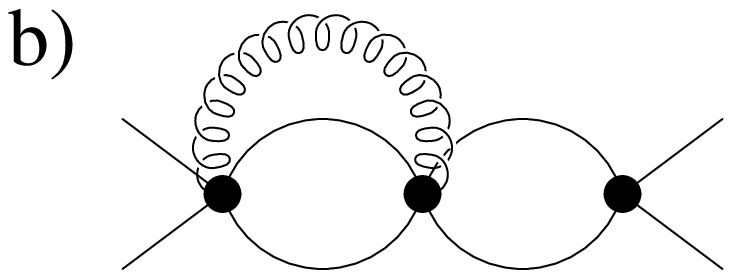}} \qquad
 \hbox{\epsfxsize=3.8cm\epsfbox{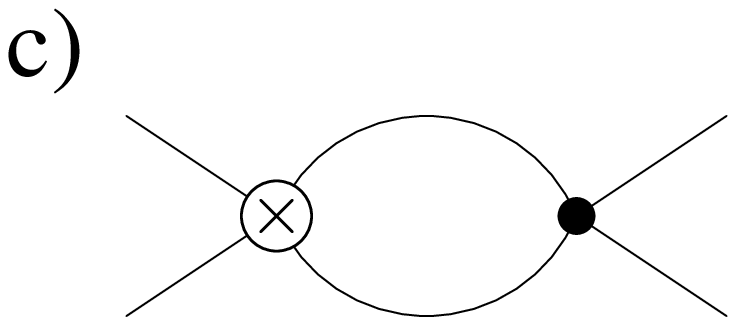}} 
 }
 \medskip\medskip
 {\tighten \caption{Graphs with an ultrasoft gluon which contributes to the 
 three-loop running of the Coulomb potential.} \label{fig_us2} }
\end{figure}

To compute the three-loop anomalous dimension for the Coulomb potential we need
to evaluate the UV divergent graphs in the effective theory that are order
$\alpha_s^4/v$. We begin by considering diagrams with an ultrasoft gluon. In
Coulomb gauge we have graphs with ${\bf p}\cdot {\bf A}/m$ vertices as well as
the coupling of ultrasoft gluons to the Coulomb potential from the
operator~\cite{amis3}
\begin{eqnarray} \label{Lpu}
  {\cal L} &=& {2 i\: \tb{ {\cal V}_c^{(T)} } \, f^{ABC}\over {\mathbf
  k}^4}\mu_S^{2\epsilon}\mu_U^\epsilon \:
    {\mathbf k}\cdot (g {\bf A}^C) \: \psip{p^\prime}^\dagger\:
  T^A {\psip p}\: \chip{-p^\prime}^\dagger\: \bar T^B {\chip {-p}}{} \,.
\end{eqnarray}
All graphs with two ${\bf p}\cdot {\bf A}$ vertices are UV finite or
are canceled by two loop graphs with insertions of the one-loop counterterms
for ${\cal V}_2$ and ${\cal V}_r$ computed in Ref.~\cite{amis}.  The remaining
diagrams are shown in Figs.~\ref{fig_us1} and \ref{fig_us2}.  Graphs
\ref{fig_us1}a through \ref{fig_us1}e have UV subdivergences which are
exactly canceled by the diagram with  ${\cal V}_k$ counterterms shown in 
Fig.~\ref{fig_us1}f.  These graphs contain subdivergences that were responsible
for the running of the $1/(m|{\bf k}|)$ potential at
two-loops~\cite{amis3}. Graph \ref{fig_us1}g is UV finite.

The divergent diagrams with an ultrasoft gluon which are not completely
canceled by a counterterm diagram are shown in Fig.~\ref{fig_us2}.
Consider the three-loop graph in Fig.~\ref{fig_us2}a with momenta $l$ for the
loop with the ultrasoft gluon and $k$ and $q$ for the remaining loops.  After
performing the $k^0$ and $q^0$ integrals by contours, the loop integration
involving $l$ is
\begin{eqnarray}
   \int d^d l \: \frac{ \delta^{ij}-{\mathbf l^i l^j/l^2}}{
  l^2 (l^0+{E}-{\bf k^2}/{m}) (l^0+{E}-{\bf q^2}/{m})    } \,.
\end{eqnarray}
From this expression we see that the ultrasoft momentum $l$ and potential
momentum, ${\bf k}$ and ${\bf q}$, are not completely separable since they
appear in the same propagator.  The $l$ integration produces an UV divergence
while the remaining integrations are UV and IR finite\footnote{\tighten For
static quarks this three-loop graph also has an IR divergence~\cite{static1},
but in the non-static case we find that this divergence is regulated by the
quark kinetic energy.}. Evaluating the remaining integrals gives
\begin{eqnarray} \label{us4}
 {\rm Fig.~\ref{fig_us2}a} &=& \frac{4i}{3} ({\cal C}_{\ref{fig_us2}a}) 
  \frac{\Big[{\cal V}_c^{(T)}(\nu)\Big]^3 \alpU(\nu)
  \mu_S^{2\epsilon}}{(4\pi)^3\: \bf k^2} \bigg[ \frac{1}{\epsilon} 
  + \ln\Big(\frac{\mu_U^2}{E^2}\Big) + 2\ln\Big(\frac{\mu_S^2}{\bf k^2}\Big) 
  + \ldots \bigg] \,,
\end{eqnarray}
where the color factor is
\begin{eqnarray} 
  ({\cal C}_{\ref{fig_us2}a}) &=& C_A C_1 \bigg[ \frac{(C_A\!+\!C_d)}{8} 
  1\otimes 1 + T\otimes\bar T \bigg] \,,
\end{eqnarray}
and for gauge group SU($N_c$), $C_d=N_c-4/N_c$ and $C_1=(N_c^2-1)/(4 N_c^2)$.
The graph in Fig.~\ref{fig_us2}c involves the iteration of a $1/{\bf k^2}$
potential and a ${\cal V}_k$ counterterm and also has a Coulombic infrared
divergence.  This graph cancels the corresponding product of IR and UV
divergences arising in Fig.~\ref{fig_us2}b. The sum of graphs in
Fig.~\ref{fig_us2}b,c still has an UV divergence, and we find
\begin{eqnarray} \label{us3}
{\rm Fig.~\ref{fig_us2}b+\ref{fig_us2}c} &=&
  -\frac{4i }{3}  ({\cal C}_{\ref{fig_us2}bc}) \frac{\Big[{\cal V}_c^{(T)}(\nu)
 \Big]^3 \alpU(\nu)\mu_S^{2\epsilon}}{(4\pi)^3\: \bf k^2} 
  \bigg[ \frac{1}{\epsilon} + \ln\Big(\frac{\mu_U^2}{E^2}\Big) 
  + 2\ln\Big(\frac{\mu_S^2}{\bf k^2}\Big) + \ldots \bigg] \,,
\end{eqnarray}
where the color factor is
\begin{eqnarray} 
  ({\cal C}_{\ref{fig_us2}bc}) &=& C_A \bigg[ \frac{C_1 (C_A\!+\!C_d)}{8} 
  1\otimes 1 - \Big( C_1 + \frac{(C_A\!+\!C_d)^2}{32} \Big) T\otimes\bar T
  \bigg] \,.
\end{eqnarray}
The sum of divergences in Eqs.~(\ref{us4}) and (\ref{us3}) are canceled by a
three-loop counterterm for ${\cal V}_c$. Differentiating with respect
to $\ln\mu_S$ and $\ln\mu_U$ gives the anomalous dimensions
\begin{eqnarray}
  \gamma_U &=&  \frac{8}{3} \bigg[ 2 C_A C_1+\frac{C_A(C_A+C_d)^2}{32} 
   \bigg] \frac{\Big[{\cal V}_c^{(T)}(\nu)\Big]^3}
  {(4\pi)^3}\: \alpU(\nu)\  \left(T\otimes\bar T\right) \,,\nn\\
  \gamma_S &=&  \frac{16}{3} \bigg[ 2 C_A C_1+\frac{C_A(C_A+C_d)^2}{32} 
   \bigg]  \frac{\Big[{\cal V}_c^{(T)}(\nu)\Big]^3}
  {(4\pi)^3}\: \alpU(\nu)\ \left(T\otimes\bar T\right) \,.
\end{eqnarray}
The total anomalous dimension from the ultrasoft diagrams is
$\gamma=2\gamma_U+\gamma_S$, so
\begin{eqnarray} \label{ad1}
  \gamma &=& \frac{32}{3} \bigg[ 2 C_A C_1+\frac{C_A(C_A+C_d)^2}{32} 
   \bigg]   \frac{\Big[{\cal V}_c^{(T)}(\nu)\Big]^3}
  {(4\pi)^3}\: \alpU(\nu)\ \left(T\otimes\bar T\right) \,.
\end{eqnarray}

\OMIT{The UV divergence in the ultrasoft graphs is matched by an IR divergence
in the purely soft diagrams at order $\alpha_s^4/v$. This IR divergence is fake;
it is not associated with $\lqcd$ physics but originates from the separation of
soft and ultrasoft gluon fields. It induces an additional purely soft anomalous
dimension (as discussed further in Appendix~\ref{UVIR}):}

The presence of an ultraviolet divergence in the ultrasoft graphs induces  
an additional ultraviolet divergence in the soft graphs (as discussed 
further in Appendix~\ref{UVIR}).  This divergence induces an additional
contribution to the soft anomalous dimension:
\begin{eqnarray} \label{ad2}
  \gamma_S &=& -{8} \bigg[ 2 C_A C_1+\frac{C_A(C_A+C_d)^2}{32} 
   \bigg]\:  \alpS^4(\nu)\  \left(T\otimes\bar T\right) \,.
\end{eqnarray}
For the Coulomb potential the remaining UV divergences in the soft graphs
correspond to divergences which are canceled in the static calculation by field,
vertex, and coupling renormalization. As discussed before, these divergences
give a contribution proportional to the three-loop $\msb$ $\beta$-function, so
for the color singlet channel we have the additional contribution
\begin{eqnarray} \label{ad3}
  \gamma_S^{(s)} &=& 2 C_F \beta_2 \frac{\alpS^4(\nu)}{(4\pi)^2} \,.
\end{eqnarray}
For QCD, $\beta_2=2857/2-5033 n_\ell/18 + 325 n_\ell^2 /54$ for $n_\ell$ light flavors.

Combining Eqs.~(\ref{ad1})--(\ref{ad3}) in the color singlet channel and using
the LL relation ${\cal V}_c^{(T)}(\nu)=4\pi\alpS(\nu)$ gives the total
anomalous dimension for the Coulomb potential to three-loop order
\begin{eqnarray} \label{adfin}
 \gamma_{\rm total}^{(s)} &=& 2 C_F \bigg[ \beta_0 \alpS^2(\nu) 
    +\beta_1 \frac{\alpS^3(\nu)}{4\pi}
    +\beta_2 \frac{\alpS^4(\nu)}{(4\pi)^2} \bigg] \nn\\
  && - {C _A^3 C_F } 
  \bigg[ \frac{4}{3} \alpS^3(\nu)\alpU(\nu)-\alpS^4(\nu) \bigg] \,.
\end{eqnarray}
Solving this equation with the two-loop boundary condition ${\cal V}_c^{(s)}(1)
= -4\pi C_F \alpha_s(m)$, the NNLL result for the running Coulomb
potential is
\begin{eqnarray} \label{Vcfin}
  {\cal V}_c^{(s)}(\nu) &=& -4\pi C_F \ams{[3]}(m\nu) 
 +\frac{8\pi C_F C_A^3}{3\beta_0} \: \alpha_s^3(m)\: 
 \bigg[ \frac{11}{4}- 2z- \frac{z^2}{2} - \frac{z^3}{4} + 4\ln(w) \bigg] \,, 
\end{eqnarray}
where 
\begin{eqnarray}
  z=\frac{ \ams{[1]}(m\nu) }{ \alpha_s(m) }\,,\qquad\quad
  w=\frac{ \ams{[1]}(m\nu^2) }{ \ams{[1]}(m\nu) } \,.
\end{eqnarray}

To compare to the static potential result from Ref.~\cite{PSstat} we can expand
the running couplings in $\alpha_s(m)$. Ignoring the $\ams{[3]}(m\nu)$ term, the
remaining logarithms in the ${\cal V}_c^{(s)}(\nu)$ coefficient are
\begin{eqnarray} \label{expVc}
  -\frac{1}{3}\, {C_F C_A^3}\,  \alpha_s^4(m) \ln(\nu) 
  +\frac{2\beta_0}{3\pi}\, {C_F C_A^3}\, \alpha_s^5(m) \ln^2(\nu) 
  -\frac{17\beta_0^2}{18\pi^2}\, {C_F C_A^3}\, \alpha_s^6(m) \ln^3(\nu) 
  +  \ldots \,.
\end{eqnarray} 
Expanding the analogous contribution to the static result in  Eq.~(\ref{psVc}) 
from Ref.~\cite{PSstat}, taking $\mu=m\nu^2$ and $r=1/m\nu$, gives
\begin{eqnarray} \label{expVstat}
 -\frac{1}{3}\, {C_F C_A^3}\,  \alpha_s^4(m) \ln(\nu) 
  +\frac{3\beta_0}{4\pi}\, {C_F C_A^3}\, \alpha_s^5(m) \ln^2(\nu) 
  -\frac{77\beta_0^2}{72\pi^2}\, {C_F C_A^3}\, \alpha_s^6(m) \ln^3(\nu) 
  +  \ldots \,.
\end{eqnarray}
From Eqs.~(\ref{expVc}) and (\ref{expVstat}) we see that the single $\ln(\nu)$
terms agree, but the higher order logarithms differ.  

\begin{figure}
 \centerline{
 \hbox{ \epsfxsize=4.2cm\epsfbox{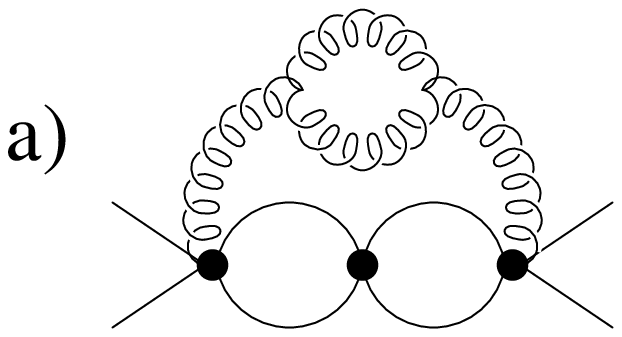}} \qquad 
 \hbox{\epsfxsize=4.2cm\epsfbox{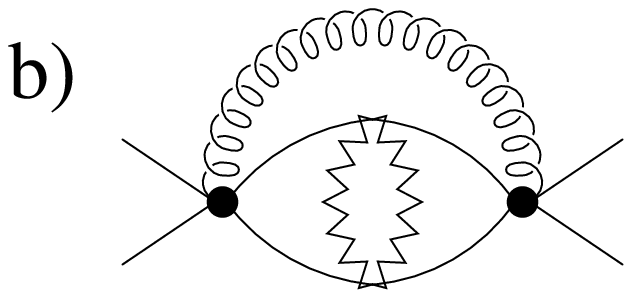}} 
 }
 \medskip\medskip
 {\tighten \caption{Graphs with double logarithms that are determined from 
Fig.~\ref{fig_us2}a.} \label{fig_usv} }
\end{figure}

The origin of the difference between the running of the Coulomb and static
potentials is the relation between scales for the case of moving versus fixed
quarks. For the Coulomb potential the non-relativistic quarks obey the
dispersion relation $E={\bf p}^2/(2m)$, which relates the energy and momentum
scales, and logarithms of $E$, ${\bf p}$ and ${\bf k}$ cannot be treated
independently. The anomalous dimension in Eq.~(\ref{adfin}) generates
$\ln^2(\nu)$ terms that reproduce the double logarithms that appear when
diagrams such as the ones in Fig.~\ref{fig_usv} are evaluated at the hard scale
$m$.  The graph in Fig.~\ref{fig_usv}a includes a vacuum polarization loop for
the ultrasoft gluon which only sees the scale $mv^2$, while the soft loop in
Fig.~\ref{fig_usv}b sees only the scale $mv$. Even though each of the two
logarithms in Fig.~\ref{fig_usv}b comes from a different ratio of low energy
scales, the modes in the graph must be included at the same time since in
the effective theory the division between soft and ultrasoft modes (and the
multipole expansion) occur right at the scale $m$. Furthermore, for all scales
below $m$ the couplings for soft and ultrasoft gluons run at different rates.
As mentioned in the introduction, this correlation of $E$, ${\bf p}$ and the
momentum transfer ${\bf k}$ has been tested successfully for bound
states in QED~\cite{amis4}. 
 
In contrast, consider the situation with two static quarks where the distance
between them, $r\sim 1/|{\bf k}|$, is held fixed externally, and the energy
fluctuations are about $E=0$. In this case the scales $r$ and $E$ are not
correlated.  Furthermore, operators with powers of $1/m$ play no role in the
calculation of the anomalous dimension, unlike the Coulombic case with an
expansion in the velocity.  The difference between the static and non-static
calculations occurs essentially because neither the $m\to \infty$ nor the $v\to
0$ limit of the effective theory is the same as the static theory.


\section{The NNLL energy for a $Q \overline Q$ bound state}\label{sec:nnll}

The energy at NNLL has contributions from matrix elements of operators of 
order
\begin{eqnarray}
 && \qquad  \Big\{ \frac{\alpha_s^4}{v^2},\ \frac{\alpha_s^3}{v},\ 
  {\alpha_s^2}v^0,\ \alpha_s v,\ v^2 \Big\}\ 
\,. \nn  
\end{eqnarray}
There are contributions from tree level matrix elements, which include the ${\bf
p^4}/m^3$ operator ($\sim v^2$), the order $v$ potentials with LL coefficients
($\sim \alpha_s v$), the order $v^0$ potentials with NLL coefficients ($\sim
\alpha_s^2 v^0$), and the correction to the energy from the Coulomb potential
with the NNLL coefficient ($\sim \alpha_s^3/v$). We also have the matrix element
of the order $\alpha_s^3/v$ one loop and two loop soft diagrams in
Fig.~\ref{eft_s2}.  Finally, there are the double insertions of two
$\alpha_s^2/v$ soft diagrams ($\sim \alpha_s^4/v^2$).  For simplicity all
$n_\ell$ light quarks are taken to be massless.

The contributions from tree level matrix elements are 
[$a_c(\nu)=-{\cal V}_c^{(s)}(\nu)/4\pi$]
\begin{eqnarray}
\Big\langle V^{(0)}\Big\rangle &=& 
  \frac{m}{4n^3(2l+1)}\:\big[\tb{ a_c(\nu) } \big]^2\:
   \tb{ {\cal V}_k^{(s)}(\nu) } \,,  \\[5pt]
\Big\langle  V^{(1)}  \Big\rangle 
  &=& \frac{m}{4n^3} \big[{\tb{ a_c(\nu) } } \big]^3 
 \Bigg\{ \frac{ \delta_{l0}}{2\pi} \:
  \Big[ \tb{ {\cal V}_2^{(s)}(\nu) } + s(s+1) \tb{ {\cal V}_s^{(s)}(\nu) } \Big]
  - \frac{ (2l+1-4n)}{8\pi n (2l+1)}\: \tb{ {\cal V}_r^{(s)}(\nu) }  \nn\\[3pt]
 &+&\frac{ 
  X_{ljs}\,\delta_{s1}\,(1-\delta_{l0})}
 {4\pi l (l+1)(2l+1)}\: \tb{ {\cal V}_\Lambda^{(s)}(\nu) } 
  + \frac{3 \langle S_{12}\rangle_{ljs}\,\delta_{s1}\,(1-\delta_{l0}) }
  {4\pi l (l+1)(2l+1)}\: \tb{ {\cal V}_t^{(s)}(\nu) } \Bigg\} \,, \nn \\[5pt]
\Bigg\langle \frac{-\bf p^4}{4 m^3} \Bigg\rangle &=&  
  \frac{m}{16n^3}\:\big[\tb{ a_c(\nu) }  \big]^4
  \:\bigg[ \frac{3}{4n}- \frac{2}{2l+1} \bigg]  \,, \nn
\end{eqnarray}
where the Wilson coefficients ${\cal V}^{(s)}_{c,k,2,s,r,\Lambda,t}$ are defined
in section~\ref{sec:pots}. The matrix elements of the soft loops are
\begin{eqnarray}
  && \left\langle\: i\!\!\!\! \begin{picture}(55,20)(-10,1)
  \epsfxsize=1.2cm \lower13pt \hbox{\epsfbox{eft_s1.eps}}
 \end{picture} \!\!\!+\ldots \right\rangle
  =   -\frac{m C_F\alpS^3(\nu)\,\tb{ a_c(\nu) } }{8\pi^2 n^2}\: \Bigg\{ 
  \Big( \frac{\beta_1}{2}+\beta_0 a_1 \Big) \Bigg( \ln\bigg(\frac{n\:\nu}
  {\tb{ a_c(\nu) }}\bigg) +\Psi(n+l+1) +\gamma_E\Bigg)   \nn\\
  && \quad\quad
  + \frac{a_2}{4} +\beta_0^2\: \Bigg( \ln^2\bigg(\frac{n\:\nu}{\tb{ a_c(\nu) }}
  \bigg) + 2 \ln\bigg(\frac{n\:\nu}{\tb{ a_c(\nu) }}\bigg)
 \Big[ \Psi(n+l+1)+\gamma_E \Big]+ N_{2}(n,l)  
  \Bigg) \Bigg\} \,, \\[7pt]
&& 
 \Bigg\langle   T\bigg\{  
i \begin{picture}(45,20)(-5,1) 
 \epsfxsize=1.0cm \lower8pt \hbox{\epsfbox{soft1.eps}} \end{picture} 
 \!\!\! ,\  
i\! \begin{picture}(45,20)(-5,1) 
 \epsfxsize=1.0cm \lower8pt \hbox{\epsfbox{soft1.eps}} \end{picture}\!\!\! 
 \bigg\}
 \Bigg\rangle 
= -\frac{m C_F^2\alpS^4(\nu)}{16\pi^2 n^2}\: \left\{\frac{a_1^2}{4}+ \beta_0 a_1 
 \Bigg( \ln\bigg(\frac{n\:\nu}{\tb{ a_c(\nu) }}\bigg) \!+\! 2 N_1(n,l) \!+\! 
 \gamma_E \Bigg) \right. \nn\\
 && \qquad \left. + \beta_0^2\Bigg(\ln^2\bigg(\frac{n\:\nu}{\tb{ a_c(\nu) }}
  \bigg) + 4\ln\bigg(\frac{n\:\nu}{\tb{ a_c(\nu) }}\bigg) 
 \Big[ N_1(n,l)+\frac{\gamma_E}{2} \Big] + 4 N_0(n,l)+4\gamma_E N_1(n,l)
 +\gamma_E^2   \Bigg)  \right\}   \,, \nn  
\end{eqnarray}
where the functions $N_{0,1,2}(n,l)$, $\langle S_{12}\rangle_{ljs}$ and
$X_{ljs}$ are obtained from Refs.~\cite{Yndurain,PY} and are summarized in
Appendix~\ref{Ns}.  The only additional contribution is the result for the NLL
energy in Eq.~(\ref{eNLL}) with the Wilson coefficients evaluated at one higher
order.

Again, at the scale $\nu=v_b$ there are no large logarithms in the matrix
elements; the large logarithms are summed up into the Wilson coefficients. For
$\nu=v_b$ the energy at NNLL order in terms of the pole mass $m$ reads
\begin{eqnarray} \label{eNNLL}
\Delta E &=& \Delta E^{LL} + \Delta E^{NLL}+ \Delta E^{NNLL} \\[2mm]
 &=&  -\frac{m}{4n^2}\: \big[ \tb{ a_c(v_b) } \big]^2 
 - \frac{m\,C_F\,\alpS^2(v_b) \tb{ a_c(v_b)} }{(8\pi\,n^2)}
 \bigg[\, 2\beta_0\Big(\psi(n+l+1) + \gamma_E \Big) + {a_1} \,\bigg] \nn\\[2mm] 
 &+& \frac{m}{4n^3(2l+1)}\:\big[\tb{ a_c(v_b)  } \big]^2\ 
   \tb{ {\cal V}_k^{(s)}(v_b) } 
 + \frac{m}{16n^3}\: \big[\tb{ a_c(v_b) }  \big]^4
 \:\bigg[ \frac{3}{4n}- \frac{2}{2l+1} \bigg]  \nn\\[1mm]
  &+& \frac{m}{4n^3} \big[\tb{ a_c(v_b) }  \big]^3 
  \Bigg\{ \frac{ \delta_{l0}}{2\pi} \:
  \Big[\tb{ {\cal V}_2^{(s)}(v_b) } + s(s+1)\tb{ {\cal V}_s^{(s)}(v_b) } \Big] -
  \frac{ (2l+1-4n)}{8\pi n (2l+1)}\: \tb{ {\cal V}_r^{(s)}(v_b) } \nn\\[3pt]
 & & +\frac{ 
  X_{ljs}\delta_{s1}(1-\delta_{l0})}
  {4\pi l (l+1)(2l+1)}\: \tb{ {\cal V}_\Lambda^{(s)}(v_b) } 
  + \frac{3 \langle S_{12}\rangle_{ljs}\delta_{s1}(1-\delta_{l0}) }
  {4\pi l (l+1)(2l+1)}\: \tb{ {\cal V}_t^{(s)}(v_b) } \Bigg\}  \nn \\[1mm]
 &-& \frac{m\,C_F\,\alpS^3(v_b)\,\tb{ a_c(v_b) }}{8\pi^2 n^2} 
 \: \Bigg\{ \,\beta_0^2\:  N_{2}(n,l) +  \bigg( \frac{\beta_1}{2}+
 \beta_0 a_1 \bigg)
\bigg[ \Psi(n+l+1)+\gamma_E \bigg]   
  + \frac{a_2}{4}\,   \Bigg\} \nn \\[1mm]
&-& \frac{m C_F^2\alpS^4(v_b)}{16\pi^2 n^2}\: \left\{ \beta_0^2\bigg[ 4 N_0(n,l) 
 +4\gamma_E N_1(n,l)+\gamma_E^2 \bigg] + \beta_0 a_1 
 \bigg[ 2 N_1(n,l)+ \gamma_E 
 \bigg] +  \frac{a_1^2}{4}\right\}   \,. \nn 
\end{eqnarray}
When this expression is expanded in powers of $\alpha_s$ at a given
renormalization scale the LL, NLL, and NNLL predictions for the energy become
series in $(\alpha_s\ln\alpha_s)$ as in Eq.~(\ref{Ell}). The terms beyond NNLO
that are determined unambiguously are those up to $m\alpha_s^4 (\alpha_s
\ln\alpha_s)^k$, $k\ge 1$.  From Eq.~(\ref{eNNLL}) we see that up to NLL the
series are determined by the running of $\ams{}(\mu)$ since up to NLL order we
have $a_c(\nu)/C_F=\alpS(\nu)=\ams{}(m\nu)$.  At NNLL order the series is no
longer just determined by the QCD $\beta$-function since operators besides the
strong coupling have non-trivial anomalous dimensions.

It is well known that the convergence of predictions for $\Delta E$ in terms of
the pole mass are plagued by the presence of infrared renormalons. If
predictions are made in terms of a short distance mass such as the $\msb$ mass
the leading renormalon in the pole mass and $1/{\bf k}^2$ potentials
cancel~\cite{renorm1,renorm2} and the convergence of the perturbation series is
improved.  A phenomenological analysis, which includes the issue of renormalon
cancellation in the presence of resummed logarithms, will be carried out
elsewhere.

Past finite order predictions for the $Q\bar Q$ energies have typically been
made using the strong coupling evaluated at the soft scale $mv$.  At LO and NLO
this is the natural choice since all logarithms $\alpha_s^2 (\alpha_s\ln\mu)^k$
and $\alpha_s^3 (\alpha_s\ln\mu)^k$ in higher order matrix elements are
minimized at $\mu=|{\bf k}|$. Effectively this choice turns the LO and NLO
results into the LL and NLL predictions.  However, at NNLO this choice of $\mu$
may not be optimal since matrix elements begin to involve factors of $\alpha_s^4
[\alpha_s\ln(\mu/E)]$.  The NNLL prediction is not generated by a simple
replacement rule and instead involves the non-trivial Wilson coefficients ${\cal
V}_{c,k,2,s,r,\Lambda,t}(\nu)$.

\begin{figure}[t] 
\begin{center}
 \leavevmode
 \epsfxsize=6cm
 \epsffile[220 580 420 710]{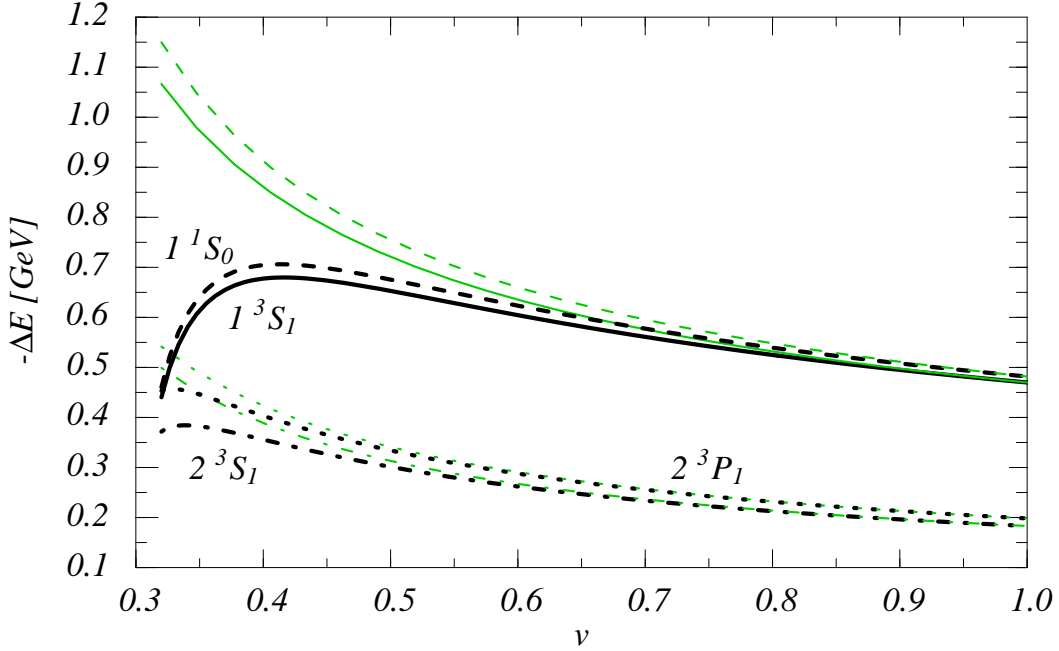}
\vskip 4.9cm 
{\tighten 
\caption{\label{fig:nnllplots} Comparison of the NNLL binding energy predictions
(thick black lines) with the NNLO predictions (thin green lines) for the
$1\,{}^3S_1$ (solid), $1\,{}^1S_0$ (dashed), $2\,{}^3S_1$ (dot-dashed), and
$2\,{}^3P_1$ (dotted) states, and for different values of the subtraction
velocity $\nu$.  } }
\end{center}
\end{figure}

In Fig.~\ref{fig:nnllplots} we compare the NNLL energy predictions at
subtraction velocity $\nu$ (thick black lines) with the NNLO predictions with
coupling $\ams{}(m\nu)$ (thin green lines), to illustrate the impact of summing
the logarithms.
The displayed states include $n\,{}^{2S+1}L_J=1\,{}^3S_1$ (solid lines),
$1\,{}^1S_0$ (dashed lines), $2\,{}^3S_1$ (dash-dotted lines), and $2\,{}^3P_1$
(dotted lines).  As input we have chosen $m_b=4.8$~GeV for the bottom quark pole
mass, $\alpha_s^{(n_\ell=4)}(m_b)=0.22$, and have included three-loop running
for the evolution of the strong coupling to lower scales.  The NNLO and the NNLL
energies are equal for $\nu=1$, because no logarithms are summed into the Wilson
coefficients. The summation has the largest impact on the $n=1$ S-wave states,
and in all cases reduces the size of the binding energy.  In
Table~\ref{table_nums} we summarize for $\nu$ of order a typical quark velocity,
$\nu=(0.35,0.4)$, values for the NNLO result and the renormalization group
improved NNLL calculation. Relative to the NNLO results the scale uncertainty in
the NNLL predictions is somewhat reduced.

The summation of NNLL logarithms also reduces the size of the ground state
hyperfine splitting, but increases the size of the analog of the QED Lamb shift,
(the $2{}^3S_1\!-\!2{}^3P_1$ splitting).  For $\nu=(0.35,0.4)$ and at NNLO the
hyperfine splitting and the $2{}^3S_1\!-\!2{}^3P_1$ splitting are
\begin{eqnarray}
  E(1{}^3S_1) - E(1{}^1S_0)=(68,52)~{\rm MeV} \,, \qquad
  E(2{}^3S_1)- E(2{}^3P_1) =(39,35)~{\rm MeV} \,,
\end{eqnarray}
while at NNLL order we find
\begin{eqnarray}
  E(1{}^3S_1) - E(1{}^1S_0)=(27,27)~{\rm MeV} \,, \qquad
  E(2{}^3S_1) - E(2{}^3P_1)=(64,47)~{\rm MeV} \,.
\end{eqnarray}
Summing the logarithms reduces the perturbative contribution to the hyperfine
splitting by a factor of two.  The $2{}^3S_1\!-\!2{}^3P_1$ splitting is fairly
sensitive to the value of $\nu$.

\begin{table}[t!]
\begin{center}
\begin{tabular}{cll|cccccc}
 & $b\bar b$ state $\: (n{}^{2S+1}L_J)$ & & $1{}^1S_0$ & $1{}^3S_1$ & $2{}^3S_1$ & $2{}^3P_1$ \\ 
 \hline
 & $\Delta E$ at NNLO (MeV) & 
  $\nu=0.35\ \ $ & $-1040$ & $-972$ & $-449$ & $-488$ \\
 & \hspace{0.9cm}\raisebox{0.2cm}{} &  
  $\nu=0.4$      & $-912$  & $-860$ & $-389$ & $-423$  \\ \hline
 & $\Delta E$ at NNLL (MeV) & 
  $\nu=0.35\ \ $ & $-641$ & $-614$ & $-382$ & $-446$ \\
 & \hspace{0.9cm}\raisebox{0.2cm}{} &  
  $\nu=0.4$      & $-705$ & $-678$ & $-356$ & $-403$ 
\end{tabular}
\end{center}
{\tighten \caption{Comparison of the NNLO and NNLL predictions for the binding
 energy corrections (in MeV) for $b\bar b$ states. The NNLL results include the
 summation of logarithms not accounted for by $\ams{}(m\nu)$. Results are
 shown for two values of $\nu$ of order the heavy quark velocity. }
\label{table_nums} }
\end{table}
\OMIT{
\begin{table}[t!]
\begin{center}
\begin{tabular}{cll|cccccc}
 & $b\bar b$ state $\: (n{}^{2S+1}L_J)$ & & $1{}^1S_0$ & $1{}^3S_1$ & $2{}^3S_1$ & $2{}^3P_1$ \\ 
 \hline
 & $\Delta E^{NNLL}-\Delta E^{NNLO}$ & 
  $\nu=0.35\ \ $ & $399$ & $358$ & $67$ & $42$ \\
 & \hspace{0.9cm}\raisebox{0.2cm}{(in MeV)} &  
  $\nu=0.4$ & $207$ & $183$ & $33$ & $20$ 
\end{tabular}
\end{center}
{\tighten \caption{Reduction in the energy (in Mev) for $b\bar b$ states from
 the summation of logarithms not accounted for by $\ams{}(m\nu)$.  The shifts
 are shown for two values of $\nu$ of order the heavy quark velocity. }
\label{table_nums} }
\end{table}
}

In order to examine the size of the logarithmic terms that are summed at NNLL
order we can expand Eq.~(\ref{eNNLL}) in powers of $\alpha_s$.  To suppress
logarithms proportional to the QCD $\beta$-function in the energy at NNLO is it
convenient to expand in the coupling $a_s \equiv\ams{} (m v_b)$:
\begin{eqnarray}\label{eNNLLexpanded}
 \Delta E = &-& \, m\,a_s^2\Big[\,\dots\,\Big]
  - m\,a_s^3\Big[\,\dots\,\Big]
  - m\,a_s^4\Big[\,\dots\,\Big]  \\[2mm]
  &-& m\,a_s^5\,\ln a_s\, \frac{C_F^2}{4\,\pi\,n^2}
  \Bigg\{\, \frac{C_A}{3}\,\bigg[\, \frac{C_A^2}{2} + 
  \frac{4\,C_A\,C_F}{n\,(2l+1)} + \frac{2\,C_F^2}{n}\,\bigg(\frac{8}{2l+1}
  -\frac{1}{n}\bigg) \,\bigg] \nn\\
& &\,\, +\, \frac{3\,\delta_{l0}\,C_F^2}{2\,n}(C_A\!+\!2C_F)
 - \frac{7\,C_A\,C_F^2\,\delta_{l0}\,\delta_{s1}}{3\,n}
 - \frac{C_A\,C_F^2\,(1\!-\!\delta_{l0})\,\delta_{s1}}
        {4\,n\,l\,(l\!+\!1)\,(2l+1)}\,\Big(\,4\,X_{ljs} 
   \!+\! \langle S_{12}\rangle_{ljs}\,\Big) \,\Bigg\} \nn\\[2mm]
 &-& m\,a_s^6\,\ln^2 a_s\,  \frac{C_F^2}{4\,\pi^2\,n^2}
\Bigg\{\,  \frac{\delta_{l0}\,C_F^2}{6\,n}\,\bigg[\,
  \beta_0\,\bigg(\frac{13\,C_A}{2}-C_F\bigg)
  +\frac{C_A}{3}\,\bigg(25\,C_A+22\,C_F\bigg)
\,\bigg] \nn\\
 & &\,\,- \frac{C_A\,C_F^2\,\delta_{l0}\,\delta_{s1}}{6\,n}\,
  \Big[\, 5\,\beta_0+ 7\,C_A \,\Big] \nn\\
 & &\,\,- \frac{C_A\,C_F^2\,(1-\delta_{l0})\,\delta_{s1}}
        {8\,n\,l\,(l+1)\,(2l+1)}\,
 \bigg[\,
   \beta_0\,\Big(\,2\,X_{ljs} 
              + \frac{1}{2}\,\langle S_{12}\rangle_{ljs}\,\Big)
   + C_A\,\Big(\,2\,X_{ljs} 
              + \langle S_{12}\rangle_{ljs}\,\Big)
 \,\bigg]
\,\Bigg\} \nn\\[2mm]
 &+& \ldots 
\,.\nn
\end{eqnarray} 
The $a_s^2$, $a_s^3$, $a_s^4$ terms are not displayed, but agree with the result
in Ref.~\cite{PY}. The terms proportional to $m\alpha_s^5\ln\alpha_s$ agree with
Ref.~\cite{a5lna}. The $m\alpha_s^6\ln^2\alpha_s$ result is new, as are higher
terms in the series.  Numerically we find
\begin{eqnarray} \label{lnseries}
  \frac{\Delta E(1{}^1S_0)}{m} &=& -0.444\, a_s^2 -1.595\, a_s^3 -9.73\, a_s^4 
     -8.56\, a_s^5\ln a_s - 3.41\, a_s^6\ln^2 a_s -15.5\, a_s^7\ln^3 a_s\nn\\
     && + \ldots \,, \nn\\
  \frac{\Delta E(1{}^3S_1)}{m} &=& -0.444\, a_s^2 -1.595\, a_s^3 -8.68\, a_s^4 
     -6.80\, a_s^5\ln a_s - 0.904\, a_s^6\ln^2 a_s -12.1\, a_s^7\ln^3 a_s\nn\\
     && + \ldots \,, \nn\\
  \frac{\Delta E(2{}^3S_1)}{m} &=& -0.111\, a_s^2 -0.546\, a_s^3 -3.07\, a_s^4 
     -0.961\, a_s^5\ln a_s - 0.113\, a_s^6\ln^2 a_s -1.73\, a_s^7\ln^3 a_s\nn\\
     && + \ldots \,, \nn\\
  \frac{\Delta E(2{}^3P_1)}{m} &=& -0.111\, a_s^2 -0.644\, a_s^3 -3.22\, a_s^4 
     -0.398\, a_s^5\ln a_s - 0.005\, a_s^6\ln^2 a_s -0.752\, a_s^7\ln^3 a_s\nn\\
     && + \ldots \,.
\end{eqnarray} 
We note that for $(1{}^1S_0,1{}^3S_1,2{}^3S_1,2{}^3P_1)$ using $a_s=0.35$, the
$a_s^5\ln a_s$ terms in Eq.~(\ref{lnseries}) give $(93\%,85\%,84\%,80\%)$ of
the complete sum of logarithmic terms.

\section{Conclusion} \label{sec:conclusion}
  
The three-loop anomalous dimension for the Coulomb potential was computed in the
presence of non-static quarks.  In terms of the subtraction velocity $\nu$ the
anomalous dimension depends on both $\alpha_s(m \nu)$ and $\alpha_s(m
\nu^2)$. Our result differs from the three-loop anomalous dimension for the
potential for static quarks.  This is due to the fact that the energy and
momentum scales are coupled for non-static quarks.

The perturbative energies for $Q\bar Q$ bound states were computed at NNLL
order, including all terms of order $m \alpha_s^4 (\alpha_s\ln\alpha_s)^k$,
$k\ge 0$.  The main effect of summing the logarithms is to reduce the binding
energy for the $1{}^1S_0$ and $1{}^3S_1$ states by an amount of order a few
hundred ${\rm MeV}$.  The effect on states with $n\ge 2$ is substantially
smaller.

\section*{Acknowledgment}

We would like to thank A.~Pineda and J.~Soto for discussions, and C.~Bauer for
comments on the manuscript.  AM and IS are supported in part by the
U.S.~Department of Energy under contract~DOE-FG03-97ER40546, and IS is also
supported in part by NSERC of Canada.


\appendix


\section{Ultraviolet and Infrared Divergences}
\label{UVIR}

In this appendix we discuss the structure of UV and IR divergences in the
effective theory and their relation to the renormalization group running. In
particular we explain why it is correct to treat all $1/\epsilon$ poles in soft
gluon loops as UV divergences. This result follows from the fact that we have
taken the full propagating gluon field and split it into two parts (soft and
ultrasoft) which fluctuate on different length scales ($mv$ and $mv^2$) and have
different Feynman rules. The information that the two gluons came from a single
field is reflected by a relation between the IR divergences in soft gluon
diagrams and the UV divergences in the ultrasoft diagrams.  Here we address how
this correlation is accounted for in the presence of the relation $\mu_U =
\mu_S^2/m$ between ultrasoft and soft subtraction scales.

For simplicity only diagrams with single $1/\epsilon$ poles and either purely
soft or ultrasoft gluons will be discussed. Also, real Coulombic IR divergences
will be dropped. A general diagram with a divergent soft gluon loop then gives a
soft amplitude with the divergence structure:
\begin{eqnarray} \label{Asoft}
 i{\cal A}^S &=& {A\over \epsilon_{UV}} + {B \over \epsilon_{IR}} +
   C \left( {1\over \epsilon_{UV}}-{1\over \epsilon_{IR}} \right)
  = {A+B \over \epsilon_{UV}} +
   (C-B) \left( {1\over \epsilon_{UV}}-{1\over \epsilon_{IR}} \right)\,.
\end{eqnarray}
Strictly speaking pure dimensional regularization does not distinguish between
UV and IR divergences; however a distinction can always be made either by
examining the analytic structure before expanding about $\epsilon=0$ or by also
performing the calculation with a different IR regulator.  The notation in
Eq.~(\ref{Asoft}) is slightly redundant so that $C$ represents diagrams
involving scaleless integrals (such as tadpole graphs), and $A$ and $B$
represent all other graphs. For example, for the Lamb shift in QED for Hydrogen
only $C$ is non-zero, while positronium also has non-zero $A$ and $B$. At the
same order in the power counting as Eq.~(\ref{Asoft}) there is an amplitude with
a divergent ultrasoft gluon loop which has the form:
\begin{eqnarray} \label{Ausoft}
 i{\cal A}^U  &=&  { (C-B) \over \epsilon_{UV}} + {D \over \epsilon_{IR}}\,.
\end{eqnarray}
The nontrivial information is that the $(C-B)$ term in Eq.~(\ref{Ausoft}) is
determined by the IR divergence in the soft amplitude in Eq.~(\ref{Asoft}). In
general $D$ is independent of $C-B$ since ultrasoft graphs are not always
proportional to $1/\epsilon_{UV} - 1/\epsilon_{IR}$. The infrared divergence in
Eq.~(\ref{Ausoft}) will match with an infrared divergence in full QCD.

The renormalization group running is determined by the UV divergences. Naively,
in Eq.~(\ref{Asoft}) the $\epsilon_{UV}$'s correspond to the scale $m$ and the
$\epsilon_{IR}$'s correspond to the scale $mv$, while in Eq.~(\ref{Ausoft}) the
$\epsilon_{UV}$'s correspond to the scale $mv$ and the $\epsilon_{IR}$'s
correspond to scales of order $mv^2$. However, examining $i{\cal A}^S+i{\cal
A}^U$ we see that the $(C-B)$ term in the soft amplitude simply pulls up or
transports the $1/\epsilon_{UV}$ in the ultrasoft graph to the hard
scale~\footnote{\tighten This addition might seem strange since it involves
canceling an UV and IR pole.  However, for a multiscale problem what is
ultraviolet and what is infrared is always relative; if we label the
$\epsilon$'s by the corresponding scale then the cancellation is just
$1/\epsilon(mv)-1/\epsilon(mv)=0$.}. The scale dependence of the coefficient
$C-B$ in Eq.~(\ref{Ausoft}) does not affect this argument since the scale
dependence of $C-B$ in Eq.~(\ref{Asoft}) can be chosen arbitrary.  Note that the
$1/\epsilon_{IR}$ divergences in soft diagrams do not always correspond to IR
divergences in QCD, which is further evidence of their unphysical nature.  This
is the case if $D\ne B-C$ such as for the two loop graphs contributing to the
running of ${\cal V}_k(\nu)$~\cite{amis3}.  Finally, we see that setting
$\epsilon_{IR} = \epsilon_{UV}$ in the soft amplitude in Eq.~(\ref{Asoft}) and
running the ultrasoft modes from $m$ to $mv^2$ with an anomalous dimension
proportional to $C-B$ and running the soft modes from $m$ to $mv$ with an
anomalous dimension proportional to $A+B$ correctly performs the running between
the scales.  This is the method used here and in Refs.~\cite{amis,amis4}.

The correspondence in Eqs.~(\ref{Asoft}) and (\ref{Ausoft}) also provides a
useful calculational tool: if the UV divergences $A+C$ in the soft diagrams are
known, and the combination $B-C$ is determined from the UV divergences in the
ultrasoft calculation, one arrives at $(A+C)+(B-C) = A+B$ which is the
combination needed to determine the soft anomalous dimension from ${\cal A}^S$.

\section{Functions that appear in $\Delta E^{\rm NNLL}$}
\label{Ns}

The following functions of the principal and orbital quantum numbers ($n,l,j$)
were derived in Refs.~\cite{Yndurain,PY} and appear in our result in
Eq.~(\ref{eNNLL}):
\begin{eqnarray}
  N_0(n,l) &=& \frac{\Psi(n+l+1)}{4}\Big[\Psi(n+l+1)-2\Big] 
  + \frac{n\Gamma(n-l)}{2\Gamma(n+l+1)} \sum_{j=0}^{n-l-2} \frac{
 \Gamma(j+2l+2)}{\Gamma(j+1)(j+l+1-n)^3}
  \nn\\
 && + \frac{n\Gamma(n+l+1)}{2\Gamma(n-l)} \sum_{j=n-l}^{\infty}  \frac{
 \Gamma(j+1)}{\Gamma(j+2l+2)(j+l+1-n)^3} \,, \\[4mm]
 N_1(n,l) &=&  \frac{\Psi(n+l+1)}{2} -\frac12 \,, \\[4mm]
 N_2(n,l) &=& \Big[ \Psi(n+l+1) +\gamma_E \Big]^2 + \Psi'(n+l+1) 
  + \frac{\pi^2}{12} \nn\\
 && +\frac{ 2 \Gamma(n-l)}{\Gamma(n+l+1)} \sum_{j=0}^{n-l-2} 
 \frac{ \Gamma(2l+2+j) }{\Gamma(j+1) (j+l+1-n)^2 }  \,,
\\[4mm]
 \langle S_{12}\rangle_{ljs} & = &
\left\{
\begin{array}{r@{\quad:\quad}l}
 \displaystyle 
 \frac{2(l+1)}{1-2l} & j=l-1 \\
 2 & j=l\\
 \displaystyle
 \frac{-2\,l}{2l+3} & j=l+1
\end{array}
\right.
\,,
\\[4mm]
X_{ljs} & = & \frac{1}{2}\Big(\,j(j+1)-l(l+1)-s(s+1)\,\Big) \,.
\end{eqnarray}

{\tighten

} 


\begin{references}

\bibitem{nrqcd}
W.E.~Caswell and G.P.~Lepage,
Phys. Lett. {\bf 167B}, 437 (1986);
G.T.~Bodwin, E.~Braaten and G.P.~Lepage,
Phys. Rev. {\bf D51}, 1125 (1995), ibid. {\bf D55}, 5853
(1997);
P.~Labelle,
Phys.\ Rev.\ D {\bf 58}, 093013 (1998);
M.~Luke and A.V.~Manohar,
Phys. Rev. {\bf D55}, 4129 (1997);
A.V.~Manohar,
Phys. Rev. {\bf D56}, 230 (1997);
B.~Grinstein and I.Z.~Rothstein,
Phys. Rev. {\bf D57}, 78 (1998);
M.~Luke and M.J.~Savage,
Phys. Rev. {\bf D57}, 413 (1998);

\bibitem{pNRQCD}
A.~Pineda and J.~Soto,
Nucl. Phys. Proc. Suppl. {\bf 64}, 428 (1998);

\bibitem{LMR} M.~Luke, A.~Manohar and I.~Rothstein,
Phys.\ Rev.\  {\bf D61}, 074025 (2000).

\bibitem{amis} A.~V.~Manohar and I.~W.~Stewart,
Phys.\ Rev.\ D {\bf 62}, 014033 (2000).


\bibitem{amis2} A.V.~Manohar and I.W.~Stewart,
Phys.\ Rev.\  {\bf D62}, 074015 (2000).

\bibitem{amis3}
A.V.~Manohar and I.W.~Stewart,
Phys.\ Rev.\ {\bf D63}, 54004 (2001).

\bibitem{PSstat}
A.~Pineda and J.~Soto,
Phys.\ Lett.\ {\bf B495}, 323 (2000).

\bibitem{amis4} A.V.~Manohar and I.W.~Stewart,
Phys.\ Rev.\ Lett.\  {\bf 85}, 2248 (2000).

\bibitem{hmst}
A.~H.~Hoang, A.~V.~Manohar, I.~W.~Stewart and T.~Teubner,
hep-ph/0011254.

\bibitem{mss1} A.V.~Manohar, J. Soto, and I.W.~Stewart,
Phys.\ Lett.\ {\bf B486}, 400 (2000).

\bibitem{Peter} M.~Peter, 
Phys.\ Rev.\ Lett.\ {\bf 78}, 602 (1997).

\bibitem{Schroeder} Y.~Schr\"oder, 
Phys.\ Lett.\ B {\bf 447}, 321 (1999).

\bibitem{chen}
Y.~Chen, Y.~Kuang and R.~J.~Oakes,
Phys.\ Rev.\  {\bf D52}, 264 (1995).

\bibitem{MW} A.~V.~Manohar and M.~B.~Wise, 
{\it Cambridge Monographs on Particle Physics, Nuclear Physics, and Cosmology,
Vol. 10}.

\bibitem{Gatheral}
J.~G.~Gatheral,
Phys.\ Lett.\ {\bf B133}, 90 (1983).

\bibitem{FrenkelTaylor}
J.~Frenkel and J.~C.~Taylor,
Nucl.\ Phys.\ {\bf B246}, 231 (1984).

\bibitem{ADM}
T.~Appelquist, M.~Dine and I.~Muzinich, 
Phys.\ Lett.\ B {\bf 69}, 231 (1977);
T.~Appelquist, M.~Dine and I.~Muzinich, 
Phys.\ Rev.\ D {\bf 17}, 2074 (1978).

\bibitem{static1}
N.~Brambilla, A.~Pineda, J.~Soto and A.~Vairo,
Phys.\ Rev.\  {\bf D60}, 091502 (1999).

\bibitem{Gries}
H.W.~Griesshammer,
Phys. Rev. {\bf D58}, 094027 (1998).

\bibitem{Fischler} W.~Fischler, 
Nucl.\ Phys.\ B {\bf 129}, 157 (1977).


\bibitem{Beneke}
M.~Beneke and V.A.~Smirnov,
Nucl. Phys. {\bf B522}, 321 (1998).

\bibitem{Yndurain}
S.~Titard and F.J.~Yndurain,
Phys.\ Rev.\ {\bf D49}, 6007 (1994).

\bibitem{PY}
A.~Pineda and F.~J.~Yndurain,
Phys.\ Rev.\ {\bf D 58}, 094022 (1998);
A.~Pineda and F.~J.~Yndurain,
Phys.\ Rev.\ {\bf D 61}, 077505 (2000).

\bibitem{renorm1}
A.~H.~Hoang, M.~C.~Smith, T.~Stelzer and S.~Willenbrock,
Phys.\ Rev.\ D {\bf 59}, 114014 (1999).

\bibitem{renorm2}
M.~Beneke,
Phys.\ Lett.\ B {\bf 434}, 115 (1998).

\bibitem{a5lna}
N.~Brambilla, A.~Pineda, J.~Soto and A.~Vairo,
Phys.\ Lett.\  {\bf B470}, 215 (1999); 
see also 
B.A.~Kniehl and A.A.~Penin, 
Nucl.\ Phys.\ {\bf B563}, 200 (1999).

\end{references}
\end{document}